\UseRawInputEncoding
\documentclass[aps,prd,preprint,superscriptaddress,nofootinbib]{revtex4}
\usepackage{amstext,amssymb}
\usepackage{amsmath}
\usepackage{graphicx}
\usepackage{slashed}
\usepackage{simpler-wick}
\usepackage{tikz}
\usepackage{tikz-feynman}
\usepackage[hyperfootnotes=true]{hyperref}
\usepackage{multirow}

\tikzfeynmanset{compat=1.1.0, warn luatex=false}

\def\nub{\bar{\nu}}
\def\M{\mathcal{M}}
\def\hc{\textup{h.c.}}

\let\vec\mathbf

\begin{document}
\title{Probing Dark Sector Particles Coupling to Neutrinos\\ with Double Beta Decay}
\author{Noor-In\`es Boudjema}
\email{noor-ines.boudjema.19@ucl.ac.uk}
\affiliation{Department of Physics \& Astronomy, UCL, London WC1E 6BT, UK}

\author{Frank F. Deppisch}
\email{f.deppisch@ucl.ac.uk}
\affiliation{Department of Physics \& Astronomy, UCL, London WC1E 6BT, UK}

\author{Antonio Herrero-Brocal}
\email{antonio.herrero@ific.uv.es}
\affiliation{Instituto de F\'ısica Corpuscular, CSIC-Universitat de Val\`{e}ncia, 46980 Paterna, Spain}

\author{Chayan Majumdar}
\email{chayanmajumdar@impcas.ac.cn}
\affiliation{Department of Physics \& Astronomy, UCL, London WC1E 6BT, UK}
\affiliation{Institute of Modern Physics, Chinese Academy of Sciences, Lanzhou 730000, China}

\author{Supriya Senapati}
\email{ssenapati@umass.edu}
\affiliation{Amherst Center for Fundamental Interactions, Department of Physics, University of Massachusetts, Amherst, MA 01003, USA}
\affiliation{Department of Applied Physics \& MIIT Key Laboratory of Semiconductor Microstructure \& Quantum Sensing, Nanjing University of Science \& Technology, Nanjing 210094, China}


\begin{abstract}
Motivated by the observation of non-zero neutrino masses and the potential for discovering physics beyond the Standard Model, numerous experiments are actively searching for neutrinoless double beta $(0\nu\beta\beta)$ decay. In all of these searches, a substantial amount of data on two-neutrino double beta $(2\nu\beta \beta)$ decay has been collected. In this work, we explore the sensitivity of current and future double beta decay experiments to a massive Majoron-like scalar particle coupled to neutrinos and potentially dark sector fermions, and compare their reach to the relevant cosmological constraints. On- and off-shell production of such a scalar leads to characteristic distortions in the emitted electron spectrum. We investigate how these distortions manifest in current and future double beta decay experiments, deriving the sensitivity to such a scenario. We project the reach of future experiments which can probe scalar-neutrino couplings of $|a_\nu| \approx 2\times 10^{-6}$ for sub-MeV scalar particles and remain sensitive to off-shell production above the $Q$-value of double beta isotopes.
\end{abstract}

\maketitle


\section{Introduction}
\label{sec:intro}
Double beta decay processes provide a sensitive window into physics beyond the Standard Model (SM). Two-neutrino double beta $(2\nu\beta\beta)$ decay, the only mode permitted within the SM and experimentally observed, is among the rarest processes ever detected, with half-lives on the order of $T_{1/2}^{2\nu} \sim 10^{19}$ years or longer \cite{Barabash:2019nnr}. In contrast, neutrinoless double beta $(0 \nu \beta \beta)$ decay, characterized by the absence of missing energy, is a clear signature of lepton number violation and provides a direct test of the Majorana nature and absolute mass scale of light neutrinos, $m_\nu$. Current experimental sensitivities reach half-lives of $T_{1/2}^{0\nu} \sim (100~\text{meV}/ m_\nu)^2 \times 10^{26}$~years, making this process a crucial probe of new physics scenarios that violate lepton number by two units \cite{Deppisch:2012nb, Graf:2018ozy, Cirigliano:2018yza, Deppisch:2020ztt}.

Although the primary goal of double beta decay experiments is the search for $0 \nu\beta\beta$ decay, they also yield precise measurements of the $2\nu\beta\beta$ decay rates and spectra across multiple isotopes. For instance, KamLAND-Zen has measured the $2\nu\beta\beta$ spectrum in $^{136}$Xe with high statistics \cite{KamLAND-Zen:2019imh}, while it is limited to the summed electron energy. In contrast, the NEMO-3 experiment, with its electron tracking capability, has enabled measurements of the individual electron energy distributions and their opening angles. This has resulted in detailed spectral data for isotopes such as $^{96}$Zr \cite{NEMO-3:2009fxe}, $^{150}$Nd \cite{NEMO-3:2016qxo}, $^{48}$Ca \cite{NEMO-3:2016mvr}, $^{82}$ Se\cite{Arnold:2018tmo} and especially $^{100}$Mo \cite{NEMO-3:2019gwo}, the latter with approximately $5\times 10^5$ recorded $2\nu\beta\beta$ decay events. These high-statistics measurements are crucial for refining theoretical inputs, such as the effective axial coupling $g_A$, which influence the interpretation of $0\nu\beta\beta$ decay limits \cite{Simkovic:2018rdz}. As the experimental exposure continues to increase, enhancing the $0\nu\beta\beta$ sensitivity, the precision of $2\nu\beta\beta$ data is also expected to improve, although the involved systematic errors and background need to improve in line with the decreased statistical uncertainty. This allows $2\nu\beta\beta$ data itself to be used to search for signals of new physics beyond the SM~(BSM). 

This may involve non-standard interactions of the SM particles involved, especially the neutrinos \cite{Deppisch:2020mxv, Deppisch:2020sqh}, modifying the decay's spectral shape or the production and emission of neutral exotic particles such as sterile neutrinos \cite{Bolton:2020ncv}, other light exotic fermions \cite{Agostini:2020cpz} and light scalars \cite{Cepedello:2018zvr}. With $2\nu\beta\beta$ $Q$ values of the order $Q \sim 2 - 4$~MeV, depending on the respective isotope, such particles can be produced if massless or lighter than a few MeV. The best-studied class of such scenarios involves the emission of a scalar particle $S$ known as the Majoron. The original Majoron model proposed a massless Goldstone boson arising from the spontaneous breaking of lepton number symmetry \cite{Chikashige:1980ui, Gelmini:1980re}, which couples to neutrinos via an interaction of the form  $g_\nu \overline{\nu^c} \nu S^*$ from which it can be produced in double beta decay as shown in Fig.~\ref{fig:feyndiag}~(left). The term \emph{Majoron} has since been generalized to refer to any electrically neutral scalar (Goldstone or otherwise) or even vector particle \cite{Burgess:1992dt, Burgess:1993xh, Carone:1993jv}. While the original models assumed a massless Majoron, more recent frameworks allow for it to be a light but massive particle \cite{Bamert:1994hb, Hirsch:1995in, Blum:2018ljv, Cepedello:2018zvr, Deppisch:2020mxv}, potentially serving as a viable dark matter candidate \cite{Berezinsky:1993fm, Garcia-Cely:2017oco, Brune:2018sab}. Current experimental searches for neutrinoless double beta decay with scalar emission $(0 \nu\beta\beta S)$ are sensitive to half-lives of the order $T_{1/2}^{0\nu S} \sim (10^{-5}/ |g_\nu|)^2 \times 10^{24}$ years \cite{KamLAND-Zen:2012uen, NEMO-3:2019gwo, Kharusi:2021jez, CUPID-0:2022yws}, assuming a massless Majoron. 

\begin{figure}[t!]
\centering
\begin{tikzpicture}
\begin{feynman}
    \vertex (b);
    \vertex[right =1.2 cm of b] (c);
    \vertex [below =1.2 cm of b] (d);
    \vertex [above =1.2 cm  of b] (a);
   \vertex [left=1.5 cm of a] (i1){$n$};
    \vertex [left=1.5 cm of d] (i2){$n$};
    \vertex[above right= 1.2 cm of a] (f1){$e^-$};
    \vertex[right =1.5 cm of a] (f2) {$p$};
    \vertex[below right= 1.2 cm of d] (f3) {$e^-$};
    \vertex[right = 1.5 cm of d] (f4){$p$};
    \diagram* {
      (i1) -- [fermion] (a),
      (i2) -- [fermion] (d),
      (a) -- [fermion] (f1),
      (a) -- [fermion] (f2),
      (d) -- [fermion] (f3),
      (d) -- [fermion] (f4),
      (a) --[majorana,edge label'=\(g_\nu\), insertion =0.5] (d),
      (b) -- [scalar,edge label'=\(S^\ast\)] (c)
    };
\end{feynman}
\end{tikzpicture}
\hspace{30pt}
\begin{tikzpicture}
\begin{feynman}
    \vertex (b);
    \vertex[right = 1.2 cm of b] (c)  ;
    \vertex [below = 1.2 cm of b] (d) ;
    \vertex [above = 1.2 cm  of b] (a) ;
    \vertex [left= 1.5 cm of a] (i1){$n$};
    \vertex [left= 1.5 cm of d] (i2){$n$};
    \vertex[above right = 1.2 cm of a] (f1){$e^-$};
    \vertex[right = 1.5 cm of a] (f2) {$p$};
    \vertex[below right = 1.2 cm of d] (f3) {$e^-$};
    \vertex[right = 1.5 cm of d] (f4){$p$};
    \vertex[below right=0.6cm and 0.8cm of c] (f5){$\chi_1$};
    \vertex[above right=0.6cm and 0.8cm  of c] (f6){$\chi_2$};

    \diagram* {
      (i1) -- [fermion] (a),
      (i2) -- [fermion] (d),
      (a) -- [fermion] (f1),
      (a) -- [fermion] (f2),
      (d) -- [fermion] (f3),
      (d) -- [fermion] (f4),
      (a) --[majorana, edge label'=\(g_\nu\),insertion =0.5] (d),
      (b) -- [scalar,edge label'=\(S^\ast\)] (c),
      (c) -- [fermion,edge label'=\(g_\chi\)] (f6),
      (c) -- [fermion] (f5),
      (c) -- [fermion] (f6)
};
\end{feynman}
\end{tikzpicture}
\caption{Feynman diagrams for double beta decay with scalar emission ($0\nu\beta\beta S$, left) and double beta decay with exotic fermion-pair emission ($2\chi\beta\beta$) via $s$-channel scalar exchange (right).}
\label{fig:feyndiag}
\end{figure}
We here focus on the modification of the double beta decay spectrum from the subsequent production of the scalar $S$ and its decay to two exotic fermions. The right diagram in Fig.~\ref{fig:feyndiag} represents such a portal to probe new fermions. In addition to this diagram, $S$ will also decay to two SM neutrinos. If $S$ can be produced on-shell, $m_S < Q$, the spectrum will be well approximated by the usual Majoron-like spectrum for electron kinetic energies $T = E_{e_1}^\text{kin} + E_{e_2}^\text{kin} < Q - m_S$. On the other hand, the finite decay width of $S$ (especially if the exotic $\chi$ particles couple strongly to $S$) and the possibility of off-shell $S$ production means that the spectrum is also modified for $T > Q - m_S$, which includes the scenario where $S$ is too heavy to be produced on-shell, $m_S > Q$. If the $\chi$ particles are massive but can still be produced on-shell, $m_{\chi_1} + m_{\chi_2} < Q$, we expect a distinctive kink in the spectrum. Searches for such exotic signatures in double beta decay are essential for advancing our understanding of neutrino properties.

Furthermore, this exotic fermion species $\chi$ can be considered as a viable dark matter (DM) candidate \cite{Huang:2014bva}, owing to its stability in presence of some discrete symmetry in the framework. The interaction between active neutrinos and such light DM, their couplings with the mediator $S$ as well as their masses can be constrained by astrophysical and cosmological bounds, i.e., coming from supernova, high energy neutrino sources; cosmological constraints arising from Cosmic Microwave Background (CMB), Big Bang Nucleosynthesis (BBN), collisional damping and present-day thermal relic density for DM particles as well as the laboratory bounds considering light meson decays to exotic particles. For a recent comprehensive overview on these constraints, we refer to \cite{Dev:2025tdv}. The possibility of inducing double beta emission by scattering of a nucleus with the DM $\chi$ background was considered in \cite{Nozzoli:2022tov}.

The paper is organized as follows. In Sec.~\ref{sec:UV}, we construct simplified models of a scalar particle coupling to neutrinos and exotic fermions. In Sec.~\ref{sec:bbdecay}, we outline the calculation of the double beta decay spectrum in the models considered. The constraints on the light $S$ and $\chi$ particles from cosmology and astrophysics are discussed in Sec.~\ref{sec:DM}, along with an overview of the most relevant laboratory-based experimental constraints. Following a discussion of our statistical approach in Sec.~\ref{sec:statistics}, we present the sensitivity of current and future experiments to select model scenarios in Sec.~\ref{sec:results} and we conclude in Sec.~\ref{sec:conclusion}.


\section{Simplified Models} 
\label{sec:UV}
In the standard $0\nu\beta\beta$ mechanism, two neutrinos annihilate due to a Majorana mass term, leaving only electrons in the final leptonic states. The existence of new BSM neutral fermions, denoted generically by $\chi$, could lead to a new variant of $\beta\beta$ decay with no neutrinos in the final state.

This hypothetical double beta decay can be first motivated by a four-fermion interaction, arising from a heavy mediator in an effective field theory (EFT) formulation. Since neutrinos produced in double beta decay arise from the SM Fermi interaction, where the leptonic part of the current is given by $J_\ell^\mu = -G_F/\sqrt{2} \bar e \gamma^\mu P_L \nu$, we need the combination $\wick{\c \nu \c \nub}$ or, equivalently, $\wick{\c \nu^c \c{\overline{\nu^c}}}$. Therefore, possible EFT operators that can give rise to the exotic double beta decay are
\begin{align}
\label{eq:Operators_EFT}
    & \mathcal{L}_{S} = 
    \frac{1}{\Lambda^2} \left[\overline{\chi_i^c} \chi_j\right] 
    \left[\overline{\nu^c} \nu\right], \nonumber\\
    & \mathcal{L}_{P} = 
    \frac{1}{\Lambda^2} \left[\overline{\chi_i^c} \gamma^5 \chi_j\right] \left[\overline{\nu^c} \gamma^5 \nu\right], \\
    & \mathcal{L}_{A} = 
    \frac{1}{\Lambda^2} \left[\overline{\chi_i^c} \gamma^5 \gamma^\mu \chi_j\right] \left[\overline{\nu^c} \gamma^5 \gamma_\mu \nu\right]. \nonumber 
\end{align}
Depending on the charge assignment, we can have only one ($i=j$) or different ($i\neq j$) generations of BSM fermion $\chi$ involved in the process. Here, $\Lambda$ corresponds to the new physics scale which is a function of the mediator mass and its coupling with $\nu$ and $\chi$. Considering a generic symmetry group $U(1)_X$ under which $\nu$ and $\chi_i$ have non-trivial charges, $Q_X^\nu$ and $Q_X^{\chi_i}$, respectively, in order to construct $U(1)_X$-invariant EFT vertices in Eq.~\eqref{eq:Operators_EFT}, we have, for example, $Q_X^\nu = -Q_X^{\chi_i} = - Q_X^{\chi_j}$.

These interactions, when probed in double beta decay experiments, could provide a promising window for discovering new physics. However, when computing the decay rate for $2\chi_S\beta\beta$ decay, some of the operators in Eq.~\eqref{eq:Operators_EFT} lead to a suppression by the small neutrino masses. In general, integrating out a mediator generates the effective operators, namely they can be realized through two distinct tree-level diagrams, via an $s$-channel or $t/u$-channel exchange. For the $s$-channel scenario with scalar mediator $S$, the Lagrangian takes the general form
\begin{equation} 
    - \mathcal{L}^s_{UV} \supset 
    \overline{\nu^c} \left(a_\nu P_L + b_\nu P_R\right) \nu S^*  
    +  \overline{\chi^c} \left(a_\chi P_L + b_\chi P_R\right)\chi S^* 
    + \hc.
\label{eq:UVcomplete_S}
\end{equation}
The resulting $S$-mediated $2\nu_S\beta\beta$ and $2\chi_S\beta\beta$ decay is not chirally suppressed by the light active neutrino masses. Moreover, the process can be dominated by the emission of $S$, i.e., scalar emission \cite{Brune:2018sab}, provided it is kinematically accessible. We can recover the purely scalar and pseudoscalar interactions by imposing $a_i = \pm b_i$, respectively. In the latter case, we reproduce the ordinary Majoron term $a_\nu \overline{\nu^c} \gamma_5 \nu S^* + \text{h.c.}$. The classical scenario with Majorana neutrinos is easily reproduced by taking $\nu = \nu^c$. In this case, $S$ must be real and the coupling includes a normalization factor of $1/2$ to avoid double-counting. Then $\frac{a_\nu}{2} \overline{\nu^c} \gamma_5 \nu S^* + \text{h.c.} = i\,\text{Im}(a_\nu) \bar\nu \gamma_5 \nu S$. For an $t/u$-channel exchange, the relevant Lagrangian is given by
\begin{equation} 
    - \mathcal{L}_{UV}^{t/u} \supset 
     \overline{\nu^c} \left(a_{\nu\chi} P_L 
    + b_{\nu\chi} P_R\right) \chi S^*
    + \hc.
\end{equation}
There is no neutrino-mass suppression of $2\chi\beta\beta$ decay in this case either, but due to the $t/u$-channel nature of the interaction, the decay amplitude acquires a nontrivial dependence on the neutrino momentum transfer $q$. This leads to integrals over $q$ with a different structure compared to standard $0\nu\beta\beta$ decay, necessitating dedicated nuclear matrix element computations.

The axial-vector operator in Eq.~\eqref{eq:Operators_EFT} arises when the mediating boson is a vector or tensor field. As before, both $s$- and $t/u$-channel completions are possible. In the $s$-channel case, a gauge boson mediates interactions between two neutrinos and two $\chi$ fermions. To reproduce the effective operator structure, the neutrino must be of Majorana type, since only then can a current-current interaction involving two neutrinos be realized. In this case, the vector interaction necessitates a chirality flip for neutrinos, leading to a suppression proportional to the neutrino mass. In contrast, the $t/u$-channel setup involves a gauge boson interacting with one neutrino and one $\chi$ fermion at the same vertex. This interaction does not require a chirality flip, and hence the resulting decay rate is not neutrino mass suppressed.

In the following, we focus on scenarios involving $s$-channel mediation by a scalar field $S$, as described in Eq.~\eqref{eq:UVcomplete_S}. In this work, the lepton number assignments of the newly introduced scalar and neutral fermion fields are chosen such that total lepton number is conserved at each interaction vertex. We work in the broken electroweak phase, thereby preserving the residual gauge symmetry group $SU(3)_C \times U(1)_\text{EM}$. Under these assumptions, neutrinos acquire Dirac masses via lepton number-conserving Yukawa interactions, and consequently, the conventional mechanism for $0\nu\beta\beta$ is absent. Furthermore, we assume that the scalar $S$ does not acquire a vacuum expectation value. However, depending on the lepton number assignments, different combinations of couplings between $S$ and the neutral fermions are possible. Therefore, in the following, we present three distinct UV-complete models, each characterized by specific lepton number assignments for the additional fields and their corresponding interaction structures. We do not attempt to explain the origin of the scalar mass, especially why it is comparatively light but we note that massive Majorons as pseudo-Goldstone boson were for example discussed in \cite{Frigerio:2011in, Garcia-Cely:2017oco}. While our frameworks are based on Dirac neutrinos, the interaction $\nu\nu S$ is equivalent to that of the ordinary Majoron and our calculations will cover the usually considered classes IA, IB and IIB of ordinary Majoron production \cite{Hirsch:1995in}, i.e., with a spectral index $n = 1$ and being dependent on the usual $0\nu\beta\beta$ decay nuclear matrix element $\mathcal{M}_{0\nu}$, as limiting cases.

For simplicity, we consider a single generation of each fermion species included, and specifically assuming the active state to be the electron neutrino. Extending to three generations of light neutrinos is straightforward. For example, allowing $S$ to couple to any combination of active states generalizes the interaction in Eq.~\ref{eq:UVcomplete_S} to $\overline{\nu^c_\alpha} \left(a_\nu^{\alpha\beta} P_L + b_\nu^{\alpha\beta} P_R\right) \nu_\beta S^*$ with $\alpha, \beta = e, \mu, \tau$. The $s$-channel production of $S$ as in Fig.~\ref{fig:feyndiag} can then be calculated by summing over all neutrino mass eigenstates but as it is electron-flavour states that are being created in the charged-current vertices, this is equivalent to assuming electron-neutrino states only, with the amplitude proportional to the $ee$ element $a_\nu^{ee}$ of the coupling matrix. 


\subsection{Model I}

We first construct a model featuring a neutral fermion sector composed of left- and right-handed neutrinos $(\nu_L^\prime, \nu_R^\prime)$ and additional singlet fermions $(\chi_L^\prime, \chi_R^\prime)$, along with a scalar singlet $S$. The fields are assigned lepton numbers as $L_{\nu_L^\prime} = +1 = L_{\nu_R^\prime}, L_{\chi_L^\prime} = +1 = L_{\chi_R^\prime}, L_S = 2$. With these charge assignments, the relevant Lagrangian takes the form 
\begin{align}
    -\mathcal{L} &= 
      m_\nu \overline{\nu_R^\prime} \nu_L^\prime + m_{\nu\chi} \overline{\chi_R^\prime} \nu_L^\prime 
    + m_{\chi\nu} \overline{\nu_R^\prime} \chi_L^\prime + M \overline{\chi_R^\prime} \chi_L^\prime 
    \nonumber \\
    &+ (a_\nu \overline{\nu^{\prime c}_L} \nu_L^\prime S^* + b_\nu \overline{\nu^{\prime c}_R} \nu_R^\prime S^* 
     + a_{\nu\chi} \overline{\nu^{\prime c}_L} \chi_L^\prime S^* + b_{\nu\chi} \overline{\nu^{\prime c}_R} \chi_R^\prime S^* 
     + a_\chi \overline{\chi^{\prime c}_L} \chi_L^\prime S^* + b_\chi \overline{\chi^{\prime c}_R} \chi_R^\prime S^*) 
     + \text{h.c.}.
\label{eq:case_I}
\end{align}

The neutral fermion mass matrix in the basis of $(\nu_L^\prime, \nu_R^{\prime c}, \chi_L^\prime, \chi_R^{\prime c})^T$ can then be written as
\begin{equation}
  \mathcal{M} =  
  \begin{pmatrix}
    0 & m_\nu & 0 & m_{\nu\chi} \\
    m_\nu & 0 & m_{\chi\nu} & 0 \\
    0 & m_{\chi\nu} & 0 & M \\
    m_{\nu\chi} & 0 & M & 0
\end{pmatrix}.
\end{equation}
However, for simplicity, assuming $m_{\chi\nu} = m_{\nu\chi}$ and considering all relevant Yukawa couplings to be real, the mass matrix $\mathcal{M}$ can be diagonalized yielding two Dirac mass eigenstates with masses,
\begin{align}
    m_\text{light,heavy} = 
    \frac{1}{2} \left|(M+m_\nu)\mp\sqrt{(M-m_\nu)^2 + 4m_{\nu\chi}^2}\right|.
\end{align}
In the hierarchical regime where $(M-m_{\nu}) \gg m_{\nu\chi}$, the masses are given by
\begin{align}
    m_\text{light} \approx \left|m_\nu - \frac{m_{\nu\chi}^2}{M-m_\nu}\right|, 
    \qquad
    m_\text{heavy} \approx M.
    \label{eq:hierarchy_I}
\end{align}

In the mass basis, the flavor eigenstates can be written as
\begin{align}
    \begin{pmatrix}
        \nu^{\prime}_{L,R} \\
        \chi^\prime_{L,R}
    \end{pmatrix} = 
    \begin{pmatrix}
         \cos\theta & \sin\theta \\
        -\sin\theta & \cos\theta
    \end{pmatrix} \cdot 
    \begin{pmatrix}
        \nu_{L,R} \\
        \chi_{L,R}
    \end{pmatrix},
\end{align}
where primed and unprimed bases correspond to interaction and mass eigenstates, respectively. These Dirac mass eigenstates are defined as, $\nu = \nu_{L} + \nu_{R}$ and $\chi = \chi_{L} + \chi_{R}$ with masses $m_\text{light}$ and $m_\text{heavy}$, respectively. In the limit of negligible light-heavy mixing $\theta \to 0$, we recover the active neutrinos and heavier exotic Dirac particles as light ($\nu_{L,R} \approx \nu^\prime_{L,R}$) and heavy ($\chi_{L,R} \approx \chi^\prime_{L,R}$) eigenstates, respectively. As an example, we can consider $M \sim 1$~MeV, $m_\text{light} \sim 10^{-2}$~eV and $m_{\nu\chi} \sim 10^2$~eV which yields a light–heavy mixing angle of $\theta \approx m_{\nu\chi} / M = 10^{-4}$.

In the mass basis, the interactions, can be written as
\begin{align}
    -\mathcal{L} &= 
      \overline{\nu^c}\lbrace  
      \left(a_\nu - a_{\nu\chi} \theta+a_\chi \theta^2\right) P_L 
    + \left(b_\nu - b_{\nu\chi} \theta+b_\chi \theta^2 \right)P_R \rbrace\nu S^* 
    \nonumber\\
    &+ \overline{\nu^c} \left \lbrace \left(a_{\nu\chi} +  
     (a_\nu-a_\chi)\theta\right) P_L + \left(b_{\nu\chi} 
    +(b_\nu-b_\chi)\theta\right) P_R \right \rbrace\chi S^* \nonumber \\
    &+ \overline{\chi^c} \lbrace  
      \left(a_\chi+a_{\nu\chi} \theta+a_\nu \theta^2 \right) P_L 
    + \left(b_\chi+b_{\nu\chi} \theta+b_\nu \theta^2 \right) P_R \rbrace \chi S^* + \text{h.c.},
\end{align}
where we have used the relation $g_\psi \overline{\psi_1^c} \psi_2 = g_\psi^T \overline{\psi_2^c} \psi_1$, following from the properties of the charge conjugation operator. We also retained the second-order mixing term $\theta^2$, otherwise ignored, accompanied by $a_{\nu, \chi}$ and $b_{\nu, \chi}$ as these couplings could enhance it. 

This above simplified model provides a minimal realization of Dirac neutrino mass through a tree-level Dirac seesaw mechanism, via mixing with an additional neutral Dirac fermion $\chi$. Here, $m_{\text{light}}$ is the light active neutrino mass and the heavier state is a sterile Dirac neutrino. Due to potentially sizeable mixing between these light and heavy states, $\chi$ is unstable and decays to three active neutrino final states, $\chi \to \nu S^\ast \to 3\nu$.


\subsection{Model I\texorpdfstring{$^\prime$}{'}}
\label{sec:modelIprime}

To ensure the stability of $\chi$ and allow it to be a viable dark matter candidate (or component of dark matter), we impose a discrete $\mathcal{Z}_2$ symmetry on Model I under which $\chi^\prime_L$ and $\chi^\prime_R$ are odd, while all other fields are even. This forbids terms that couple $\nu$ to $\chi$, leading to the Lagrangian
\begin{align}
    -\mathcal{L} &=
      m_\nu \overline{\nu_L^\prime} \nu^\prime_R 
    + M \overline{\chi^\prime_L} \chi^\prime_R 
    + (a_\nu \overline{\nu^{\prime c}_L} \nu^\prime_L 
    + b_\nu \overline{\nu^{\prime c}_R} \nu^\prime_R 
    + a_\chi \overline{\chi^{\prime c}_L} \chi^\prime_L 
    + b_\chi \overline{\chi^{\prime c}_R} \chi^\prime_R) S^*
    + \text{h.c.}.
\end{align}
The masses are already diagonal in this case and there is no mass mixing between $\nu$ and $\chi$. This then simply yields two Dirac pairs with masses $m_\text{light} = m_\nu$ and $m_\text{heavy} = M$. The interaction Lagrangian in this basis becomes 
\begin{align}
    -\mathcal{L} &=
    \overline{\nu^c} \left(a_\nu P_L + b_\nu P_R \right) \nu S^* 
    + \overline{\chi^c} \left(a_\chi P_L + b_\chi P_R \right) \chi S^* + \text{h.c.}.
\label{eq:modelI_mass_Z2}
\end{align}
As a consequence, the $\mathcal{Z}_2$-odd particle $\chi$ is stable.


\subsection{Model II}

Finally, we consider a scenario involving two distinct, exotic neutral Dirac fermions denoted as $\chi^\prime_1$ and $\chi_2$, with the lepton number assignments $L_{\nu^\prime_L} = +1 = L_{\nu^\prime_R}, L_{\chi^\prime_{1L}} = -2 = L_{\chi^\prime_{1R}}, L_{\chi_{2}} = 0, L_{S} = 2$. The active SM neutrinos retain their canonical lepton number values. The Lagrangian respecting these assignments is given by
\begin{align}
    -\mathcal{L} &= 
       m_\nu \overline{\nu_L^\prime} \nu^\prime_R + M_1 \overline{\chi^\prime_{1L}} \chi^\prime_{1R} 
     + M_2 \overline{\chi_2^{c}} \chi_2  \nonumber \\
    &+ (a_\nu \overline{\nu^{\prime c}_L} \nu_L^\prime + b_\nu \overline{\nu^{\prime c}_R} \nu^\prime_R 
     + a_{\chi_1\chi_2} \overline{\chi^\prime_{1R}} \chi_{2} + b_{\chi_1\chi_2} \overline{\chi^\prime_{1L}} \chi_{2}) S^* 
     + \text{h.c.}.
\end{align}
The corresponding neutral fermion mass matrix in the basis of $(\nu^\prime_L, \nu_R^{\prime c}, \chi^\prime_{1L}, \chi_{1R}^{\prime c}, \chi_{2})^T$ takes the form
\begin{equation} 
    \mathcal{M} \sim 
    \begin{pmatrix}
        0 & m_\nu & 0 & 0 & 0 \\
        m_\nu & 0 & 0 & 0 & 0 \\
        0 & 0 & 0 & M_1 & 0 \\
        0 & 0 & M_1 & 0 & 0 \\
        0 & 0 & 0 & 0 & M_2
    \end{pmatrix}.
\end{equation}

In this framework, we have three mass-decoupled neutral fermions: the active neutrinos form a Dirac state $\nu^\prime = \nu = \nu^\prime_L + \nu^\prime_R$ with mass $m_\nu$, the field $\chi^\prime_1 = \chi_1 = \chi^\prime_{1L} + \chi^\prime_{1R}$ becomes a Dirac state with mass $m_{\chi_1} = M_1$ and $\chi_2$ is a Majorana fermion with mass $M_2$. In the mass basis, the interactions are
\begin{align}
\label{eq:model2-massbasis}
    -\mathcal{L} &=
      \overline{\nu^c} \lbrace a_\nu P_L + b_\nu P_R \rbrace \nu S^* 
    + \overline{\chi_1} \lbrace a_{\chi_1 \chi_2} P_L + b_{\chi_1 \chi_2} P_R \rbrace \chi_2 S^* 
    + \text{h.c.}.
\end{align}
This framework is mainly intended to highlight that interactions involving both Dirac and Majorana neutral fermions through the scalar mediator are possible. 


\subsection{Scalar Decay Width}

Before discussing double beta decay in detail, we determine the total decay width of the scalar $S$ in all three models. Considering $\phi$ and $\psi_{1,2}$ as generic scalar and fermions, respectively, with the coupling $\overline{\psi^c_1} (a P_L + b P_R) \psi_2 \phi^*$ + h.c., the corresponding partial decay width of $\phi\to\psi_1\psi_2$ is given by
\begin{align}
    \Gamma(\phi\to\psi_1\psi_2) = 
    \frac{m_\phi}{16\pi} 
    & \left \{ \left(|a|^2 + |b|^2\right)\left(1 - x_1^2 - x_2^2\right) 
    - 4\text{Re}\left[a^* b\right] x_1 x_2 \right \} \nonumber \\
    & \times \sqrt{\left[1-(x_1 -x_2)^2\right]\left[1-(x_1 + x_2)^2\right]} ,
\end{align}
with the mass ratios $x_i = m_{\psi_i} / m_\phi$ ($i=1,2$). The decay width to a pair of identical fermions $\psi_1 = \psi_2 = \psi$ includes a symmetry factor and can be written as
\begin{align}
    \Gamma(\phi\to\psi\psi) = 
    \frac{m_\phi}{32\pi}
    &\left\{\left(|a|^2 + |b|^2\right)\left(1 - 2 x^2\right) 
    - 4\,\text{Re}\left[a^* b\right]x^2\right\} 
    \sqrt{1 - 4x^2},
\end{align}
with $x = m_\psi/m_\phi$. Therefore, the total decay widths for the scalar $S$ in our models are
\begin{itemize}
    \item Model I:
        \begin{equation}
            \Gamma_S = 
             \Gamma(S\to\nu\nu) + \Gamma(S\to\nu\chi) +  \Gamma(S\to\chi\chi). 
        \label{eq:modelI}
        \end{equation}
        Furthermore, the heavier state $\chi$ is not stable as it can decay via an off-shell $S$, $\chi\to\nu S^*\to 3\nu$ as well as through the SM currents due to the mixing.
        
    \item Model I$^\prime$:
        \begin{equation}
            \Gamma_S = \Gamma(S\to\nu\nu) +  \Gamma(S\to\chi\chi). 
        \label{eq:modelIp}
        \end{equation}
        Due to the imposed $\mathcal{Z}_2$ symmetry, there is no $\chi\nu S$ interaction and $\chi$ is stable.

    \item Model II:
        \begin{equation}
            \Gamma_S = 
               \Gamma(S\to\nu\nu) 
            + \Gamma(S\to\bar{\chi}_1 \chi_2).
        \label{eq:modelII}
        \end{equation}
        In this model, the heavier fermion state, say, $\chi_2$ can decay via an off-shell $S$, $\chi_2 \to \chi_1 S^* \to \chi_1 + 2\nu$. 
\end{itemize}
We note that the decay widths can be further enhanced by allowing $S$ to couple to more than one neutrino generation, including flavour-violating modes, $S\to \nu_\alpha \nu_\beta$ ($\alpha = e, \mu, \tau$). 

\begin{figure}[t!]
    \centering
    \includegraphics[width=0.8\linewidth]{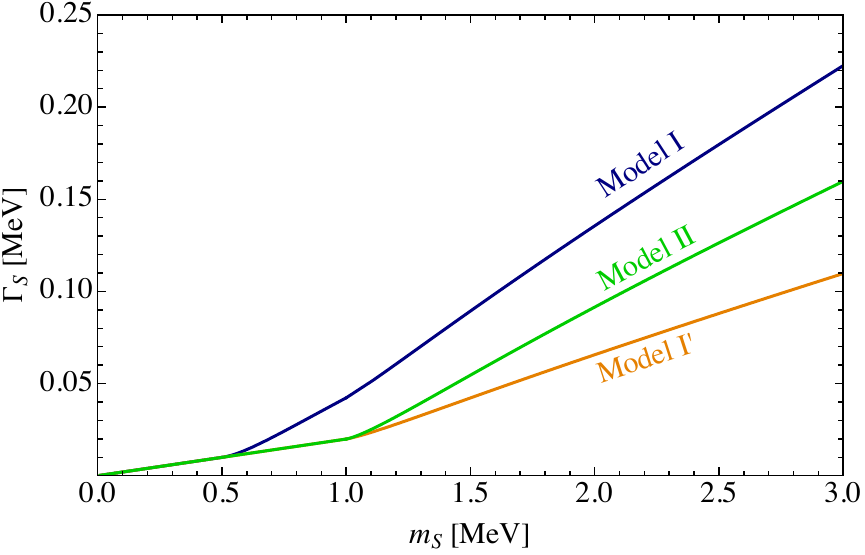}
    \caption{Total decay width of the scalar particle $S$ as a function of its mass $m_S$ in the three different models considered. The active neutrino is considered massless whereas we take $m_{\chi_{1,2}} = 0.5$~MeV. In each model, all $a_i$ and $b_i$ couplings are set to unity. The active-sterile mixing in Model~I is set to $\theta = 10^{-4}$.} 
    \label{fig:decay_width}
\end{figure}
In Fig.~\ref{fig:decay_width}, we plot the total decay widths of $S$ as a function of its mass $m_S$ in the three different models considered. Here, the active neutrino is considered massless whereas we take $m_{\chi_{1,2}} = 0.5$~MeV for the heavier fermion(s). The active-sterile mixing in Model I is set at $\theta = 10^{-4}$ and all the couplings are $a_{i} = b_{i} = 1$ ($i = \nu, \chi, \nu\chi, \chi_1\chi_2$ as discussed for each model). As expected for an MeV-scale scalar, the widths are small, reaching at most $\Gamma_S \sim 220$~keV in Model~I due its higher number of decay channels with thresholds at $m_S = 0.5$~MeV ($\nu\chi$ accessible) and 1~MeV ($\chi\chi$ accessible). The total decay width of $S$ for Model I$^\prime$ and Model~II are equal for $m_S < 1$~MeV (due to only accessible $\nu\nu$ channel). For larger $S$ masses, the Model~II decay becomes larger as evident from the Eqs.~\eqref{eq:modelIp} and \eqref{eq:modelII}.


\section{Double Beta Decay} 
\label{sec:bbdecay}

In Sec.~\ref{sec:UV}, we saw that the presence of BSM neutral fermions could lead to new phenomenology in double beta decay experiments. In particular, Sec.~\ref{sec:UV} revealed that the most promising scenario involves a scalar mediator in an $s$-channel. Let us now analyze this possibility. We will work with two different $\chi$ particles and with different couplings for the right and left-handed fermions for the sake of generality.

From Fig.~\ref{fig:feyndiag}, it can be observed that the nuclear structure of both processes ($\chi_1\chi_2\beta\beta$ decay and scalar emission) shares the same nuclear structure as the $0\nu\beta\beta$ decay, since the corresponding Feynman diagrams are identical except for the emission of a scalar particle. Consequently, all the constants involved in these processes are those appearing in $0\nu\beta\beta$ decay, with the addition of the new interaction terms. We therefore denote these common constants generically as $\mathcal{N}_{0\nu}$, which we match to the conventions used in the literature in order to ensure the correct normalization of the nuclear matrix elements (NMEs). We here present the derivation of the decay width for the generic $\chi_1\chi_2\beta\beta$ decay via an $s$-channel scalar. We also show the results for the interference with the standard $2\nu\beta\beta$ decay, in the case $S\to \bar\nu\bar\nu$, i.e., for the emission of two active anti-neutrinos. For comparison and later analysis we also show the rates for scalar particle emission $0\nu\beta\beta S$ (with ordinary Majoron emission $0\nu\beta\beta J$ as massless limit) and the SM $2\nu\beta\beta$ decay. Here, we note that we omit the calculation of $2\nu_{(t-\text{channel}~S)}\beta\beta$ via the exchange of a $t$-channel scalar as we are here mainly interested in the modification of the spectrum near the scalar production threshold and this contribution is expected to be sub-dominant due to the higher momentum $|q^2| \approx (100~\text{MeV})^2$ in the propagator \cite{Deppisch:2020sqh}. In our specific models introduced in Sec.~\ref{sec:UV}, $S$ carries a conserved lepton number charge, and the $t$-channel contribution is not allowed anyway. 


\subsection{\texorpdfstring{$\chi_1\chi_2\beta\beta$ Decay}{chi1-chi2-beta-beta Decay}}

The effective Lagrangian inducing $\chi_1 \chi_2 \beta\beta$ decay via $s$-channel exchange of a massive scalar $S$ is given by  
\begin{align}
    \mathcal{L} &=
    - \overline{\nu^c} \left(a_\nu P_L + b_\nu P_R\right) \nu S^*  
    - \overline{\chi^c_i} \left(a_\chi P_L + b_\chi P_R\right) \chi_j S^* 
    + \text{h.c.}. 
\end{align}
Here, $\nu$ is the electron-neutrino, assumed to be the only neutrino species. We will treat $\chi_i$ and $\chi_j$ as different particles in our calculations. If they are identical particles, a symmetry factor of $1/2$ must be included.

The interactions above lead to two different contributions depending on whether $S$ is considered to be a final state particle. The corresponding Feynman diagrams are shown in Fig.~\ref{fig:feyndiag}. The left diagram represents $0\nu\beta\beta S$ decay, which is allowed as long as the production of the scalar mediator $S$ is kinematically allowed, i.e., $m_S < Q$. The right diagram represents a potential portal to probe new fermions, either arising from the on-shell production of $S$ and its subsequent decay, or through an off-shell $S$. We compute the amplitude for the latter process $\chi_1 \chi_2 \beta\beta$ and its leptonic part is given by
\begin{align}
    L^{\mu\nu}_{2\chi} = 
    \frac{\mathcal{N}_{0 \nu}}{p^2 - m_S^2 + i m_S \Gamma_S}
   & \left[\bar{u}(p_1) \gamma^\mu P_L \frac{\slashed{q} + m_\nu}{q^2 - m_\nu^2} 
           \left(a_\nu^* P_R + b_\nu^* P_L\right)
          \frac{\slashed{p}-\slashed{q} + m_\nu}{(p-q)^2 - m_\nu^2} \gamma^\nu P_R v(p_2)
    \right]  \nonumber\\
    \times &\left[\bar{u}(p_3) \left(a_\chi^* P_R + b_\chi^* P_L\right) v(p_4)\right].
\end{align}
The outgoing electrons have momenta $p_1$ and $p_2$ whereas the exotic fermions $\chi_1$ and $\chi_2$ have momenta $p_3$ and $p_4$, respectively. The neutrino mass is $m_\nu$, and $m_S$ and $\Gamma_S$ are the mass and total decay width for the scalar mediator $S$, respectively, which has momentum $p = p_3 + p_4$. Due to the small $Q \sim$~MeV value in double beta decay isotopes and the tiny neutrino masses $m_\nu \lesssim 0.01$~eV, we can assume that the final state lepton momenta and the neutrino masses are negligible compared to the momentum exchanged through the neutrino propagators, $|q^2| \approx (100~\text{MeV})^2 \gg |p^2|, m_\nu^2$, and the amplitude simplifies to
\begin{equation}
    L^{\mu\nu}_{2\chi} = 
    \frac{\mathcal{N}_{0 \nu} a_\nu^*}{q^2} \frac{g^{\mu\nu}}{p^2 - m_S^2 + i m_S \Gamma_S} 
    \left[\bar{u}(p_1) P_R v(p_2)\right]
    \left[\bar{u}(p_3)\left(a_\chi^* P_R + b_\chi^* P_L\right) v(p_4)\right].
\end{equation}
Here, we have used the symmetry under the $\mu \leftrightarrow \nu$ exchange of the hadronic contribution. The factor $N_{0\nu}$ captures the overall normalization of the amplitude to recover the usual definition of the NME used in the literature. It will be resolved below as further discussed in Appendix~\ref{App:normalization}.

To compute the leptonic phase space of the process, we work in $s$-wave approximation for the outgoing electrons. For each electron, we include the Fermi function $F_0(Z_f, E_i)$ ($i = 1, 2$), which depends on the energy of the electron. It describes the modification of the electron wave function by the electromagnetic potential of the final state nucleus with charge $Z_f = Z + 2$ \cite{Doi:1985dx},
\begin{equation}
    F_0(Z_f, E) = 
    4\frac{(2 R \sqrt{E^2 - m_e^2})^{2 (\gamma_0 -1)}}{\Gamma^2(1 + 2\gamma_0)} 
    e^{\pi y} \left|\Gamma(\gamma_0 + i y)\right|^2.
\end{equation}
Here, $R = 1.2~\text{fm}\cdot A^{1/3}$ denotes the nuclear radius, $\Gamma(x)$ is the Gamma function, $\gamma_0 = \sqrt{1 - (\alpha Z_f)^2}$, and $y = \alpha Z_f E/\sqrt{E^2 - m_e^2}$, where $\alpha = 1/137$ is the fine-structure constant. As in $0\nu\beta \beta$ decay, we assume the closure approximation, which allows us to factorize the amplitude of the process into separate leptonic and hadronic contributions. The squared amplitude can then be expressed as
\begin{align}
    |i M_{2\chi}|^2 &= 
    \left|\frac{\mathcal{N}_{0\nu} a_\nu}{p^2 - m_S^2 
    + i m_S\Gamma_S}\right|^2 
    F_0(p_1) F_0(p_2) |\M_{0\nu}|^2 \nonumber\\
    &\times 4 p_1 \cdot p_2 
    \left[\left(|a_\chi|^2 +|b_\chi|^2\right) p_3 \cdot p_4 
          - 2\text{Re}\left[(a_\chi)^* b_\chi\right]m_{\chi_1} m_{\chi_2}
    \right],
\end{align}
where the NME $\M_{0\nu}$ is identical to the one in standard $0\nu\beta\beta$ decay, due to the same hadronic structure. We will write the total $\chi_1\chi_2\beta\beta$ decay width as
\begin{align}
    \Gamma_{2\chi} = 
     \mathcal{G}_{2\chi} \, \left|\M_{0\nu}\right|^2,
\end{align}
The phase space factor $\mathcal{G}_{2\chi}$ is calculated by integrating 
\begin{align}
    \mathcal{G}_{2\chi} &= \left|\mathcal{N}_{0 \nu} \right|^2 |a_\nu|^2
    \nonumber\\
    &\times\int 
    \prod_{i=1}^4 \frac{d^3 \vec{p}_i}{(2\pi)^3 2E_i} 
    4 p_1\cdot p_2 F_0(p_1) F_0(p_2)
    \frac{\left(|a_\chi|^2 + |b_\chi|^2\right) p_3\cdot p_4 
    - 2\text{Re}\left[a_\chi^* \, b_\chi\right]  m_{\chi_1} m_{\chi_2}}{\left|p^2 - m_S^2 + i m_S \Gamma_S \right|^2} \nonumber\\
    &\times \delta \left(Q - T - E_3 - E_4 \right),
\end{align}
over the phase space of the final state fermions with $p_i = (E_i, \vec{p}_i)$. First, we integrate over the phase space of the invisible fermions $\chi_1$ and $\chi_2$, yielding two integrals
\begin{align}
    I_1  &= 
    \int \prod_{i=3}^4 \frac{d^3 \vec{p}_i}{(2\pi)^3 2E_i} 
    \frac{p_3\cdot p_4}{\left|\left(p_3+p_4\right)^2 - m_S^2 + i m_S \Gamma_S\right|^2} \delta\left(Q - T - E_3 - E_4\right), \nonumber \\
    I_2 &= 
    m_{\chi_1} m_{\chi_2} \int\prod_{i=3}^4\frac{d^3\vec{p}_i}{(2\pi)^3 2E_i} 
    \frac{1}{\left|\left(p_3 + p_4\right)^2 - m_S^2 + i m_S \Gamma_S\right|^2} 
    \delta\left(Q - T - E_3 - E_4\right),
\end{align}
with the total kinetic energy of the electrons $T = E_1 + E_2 - 2m_e$. Defining  
\begin{align}
    x = \frac{2|\vec{p}_3| |\vec{p}_4|}{m_S^2 - m_{\chi_1}^2 - m_{\chi_2}^2 - 2 E_3 E_4}, 
\end{align}  
with $|\vec{p}_3| = (E_3^2 - m_{\chi_1}^2)^{1/2}$, $|\vec{p}_4| = (E_4^2 - m_{\chi_2}^2)^{1/2}$ and $E_4 = Q - T - E_3$, we can integrate analytically over $E_4$ and the angles of $\vec{p}_3$ and $\vec{p}_4$, to express the expressions as
\begin{align} 
\label{eq:I1_scalar}
    I_1(T) = 
    \frac{1}{4(2\pi)^4} \int_{m_{\chi_1}}^{Q - T - m_{\chi_2}} dE_3
    &\left\{
    \frac{m_S^2 - m_{\chi_1}^2 - m_{\chi_2}^2}{2m_S\Gamma_S} 
    \left[\arctan\zeta_+(E_3) + \arctan\zeta_-(E_3)\right]\right. \nonumber\\
    &+\left. \frac{1}{4} \log\left[\frac{\zeta_-^2(E_3) + 1}
    {\zeta_+^2(E_3) + 1} \right]
    \right\},
\end{align}
and
\begin{align} 
\label{eq:I2_scalar}
    I_2(T) = 
    \frac{1}{4(2\pi)^4}\frac{m_{\chi_1} m_{\chi_2}}{m_S \Gamma_S}
    \int_{m_{\chi_1}}^{Q - T - m_{\chi_2}}dE_3 
    \left[\arctan\zeta_+(E_3) + \arctan\zeta_-(E_3)\right].
\end{align}
Here, we have defined
\begin{align}
    \zeta_\pm(E_3) = 
    \frac{2|\vec{p}_3||\vec{p}_4|(x \pm 1)}{m_S \Gamma_S x}.
\end{align}
In Eqs.~\eqref{eq:I1_scalar} and \eqref{eq:I2_scalar}, we expressed the integrands using the relation 
\begin{align}
    \arctan\zeta_+ + \arctan\zeta_- = 
    \arctan\left(\frac{\zeta_+ + \zeta_-}{1 - \zeta_+ \zeta_-}\right),
\end{align}
as this turns out to be more numerically stable for physical, i.e., small values of $\Gamma_S$ when evaluating the integrals.

Then, $\mathcal{G}_{2\chi}$ can be expressed as
\begin{align}
    \mathcal{G}_{2\chi} &= \left|\mathcal{N}_{0 \nu} \right|^2 |a_\nu|^2
    \nonumber\\
    &\times\int\prod_{i=1}^2 \frac{d^3 \vec{p}_i}{(2\pi)^3 2E_i}
    4 p_1 \cdot p_2 F_0(p_1) F_0(p_2) 
    \left[\left(|a_\chi|^2 + |b_\chi|^2\right) I_1(T) 
    - 2\text{Re}\left[a_\chi^* b_\chi\right] I_2(T)\right].
\end{align}
It is convenient to transform this expression into one that depends on the electron kinetic energies, $T_1 = E_1 - m_e$ and $T_2 = E_2 - m_e$, and the angle $\theta_{12}$ between the emitted electrons, with $\vec{p}_1 \cdot \vec{p}_2 = |\vec{p}_1||\vec{p}_2|\cos \theta_{12}$. The fully differential decay rate can then be written as
\begin{align}
\label{eq:FullyDifferential}
  \frac{d\Gamma^{2\chi}_S}{dT_1 dT_2 d\cos\theta_{12}} &= 
  \kappa_{0\nu} \left(\frac{m_e}{2R}\right)^2 
  \left|\M_{0\nu}\right|^2 |\vec{p_1}| |\vec{p_2}| E_1 E_2 F_0(p_1) F_0(p_2) 
  \left(1 - \frac{| \vec{p}_1||\vec{p}_2|}{E_1 E_2}\cos\theta_{12}\right) 
  \nonumber\\ 
  &\times |a_\nu|^2 \left\{\left(|a_\chi|^2 + |b_\chi|^2\right) I_1(T_1+T_2) 
  - 2\text{Re}\left[a_\chi^* b_\chi\right] I_2(T_1+T_2)\right\}, 
\end{align}
where we have explicitly extracted the factor $m_e/(2R)$ to emphasize that $\M_{0\nu}$  is dimensionless and that the phase space factor $\mathcal{G}_{2\chi}$ has units of $\text{yr}^{-1}$, to conform to the usual conventions employed in $0\nu\beta\beta$ decay calculations. Here $|\vec{p}_i|$ and $E_i$ are understood to be expressed in terms of the $T_i$ and the factor $\kappa_{0\nu}$ has been obtained using the correct normalization for~$\mathcal{N}_{0\nu}$,
\begin{align}
\label{Tidep}
    \kappa_{0\nu} = \frac{\left|\mathcal{N}_{0 \nu} \right|^2 R^2}{2 \pi^4 m_e^2} = \frac{G_F^4\cos^4\theta_C}{ 8 \pi^5 m_e^2},
\end{align}
with the Cabibbo angle $\theta_C$.

Most double beta decay experiments measure only the differential decay rate $d\Gamma/dT$ with respect to the total electron kinetic energy $T = T_1 + T_2$. Integrating Eq.~\eqref{eq:FullyDifferential} over the angle and energy difference $\Delta T = T_1 - T_2$ yields the result
\begin{align}
\label{eq:2chibb}
    \frac{d\Gamma^{2\chi}_S}{dT} &= 
    \kappa_{0\nu} \left(\frac{m_e}{2R}\right)^2 \left|\M_{0\nu}\right|^2 
    |a_\nu|^2 \left\{\left(|a_\chi|^2 + |b_\chi|^2\right) I_1 (T) 
    - 2\text{Re}\left[a_\chi^* b_\chi\right] I_2(T) \right\}
    g(T) \nonumber\\
    &\times\Theta(Q - m_{\chi_1} - m_{\chi_2} - T),
\end{align}
with the Heaviside function $\Theta(x)$ and
\begin{align}
\label{Tdep}
    g(T) = 
    \int_{-T}^T d\Delta T 
    |\vec{p}_1| |\vec{p}_2| E_1 E_2 F_0(|\vec{p}_1|) F_0(| \vec{p}_2|),
\end{align}
where we express the integrand as $E_1 = \frac{1}{2}(T + \Delta T) + m_e$, $E_2 = \frac{1}{2}(T - \Delta T) + m_e$ and $|\vec{p}_i| = (E_i^2 - m_e^2)^{1/2}$ ($i=1,2$). Finally, to compute the total decay width, we integrate over~$T$,
\begin{align}
\label{eq:TotalWidth}
    \Gamma^{2\chi}_S = 
    \int_0^{Q - m_{\chi_1} - m_{\chi_2}} dT \frac{d\Gamma_{2\chi}}{dT}.
\end{align}
The above generic calculation for distinct, massive $\chi_1$, $\chi_2$ is applicable to all contributions considered here, with an additional symmetry factor of $1/2$ for identical particles, i.e., $2\nu_S\beta\beta$ and $2\chi_S\beta\beta$. The above expressions can be simplified for numerical evaluation if one or both $\chi$ are massless, due to $|\mathbf{p}_i| = E_i$.

\paragraph*{Heavy Scalar Mediator}
In the case of a heavy $s$-channel mediator, $m_S \gg Q$, equivalent to a four-fermion contact interaction $(\overline{\nu}\nu^c)(\overline{\chi^c}\chi)$, the above expressions can be further simplified,
\begin{align}
\label{eq:2chibb-heavy}
    \frac{d\Gamma^{2\chi}_\text{EFT}}{dT} &= 
    \kappa_{0\nu} 
    \left(\frac{m_e}{2R}\right)^2 
    \left|\M_{0\nu}\right|^2 
    \left\{\left(|\Lambda_a^{-2}|^2 + |\Lambda_b^{-2}|^2\right) I_1 (T) 
    - 2\text{Re}\left[(\Lambda_a^{-2})^* \Lambda_b^{-2}\right] I_2(T) \right\}
    g(T) \nonumber\\
    &\times\Theta(Q - m_{\chi_1} - m_{\chi_2} - T),
\end{align}
with the integrals
\begin{align} 
    I_1(T) &= 
    \frac{1}{(2\pi)^4}\int_{m_{\chi_1}}^{Q - T - m_{\chi_2}} 
    dE_3 E_3 (Q-T-E_3) \sqrt{E_3^2 -m_{\chi_1}^2}\sqrt{(Q-T-E_3)^2 - m_{\chi_2}^2},
\label{eq:I1_Eff}\\
    I_2(T) &= 
    \frac{m_{\chi_1}m_{\chi_2}}{(2\pi)^4}\int_{m_{\chi_1}}^{Q - T - m_{\chi_2}} 
    dE_3 \sqrt{E_3^2 -m_{\chi_1}^2}\sqrt{(Q-T-E_3)^2 -m_{\chi_2}^2}.
\label{eq:I2_Eff}
\end{align}
Equation~\eqref{eq:2chibb-heavy} is written in terms of the effective operator scales $\Lambda_a$ and $\Lambda_b$, defined by 
\begin{align}
    \frac{1}{\Lambda_a^2} = \frac{a_\nu a_\chi}{m_S^2}, \qquad
    \frac{1}{\Lambda_b^2} = \frac{a_\nu b_\chi}{m_S^2},
\end{align}
in the case of $s$-channel scalar mediation considered here.


\subsection{\texorpdfstring{Interference between SM $2\nu\beta\beta$ and Scalar-induced $2\nu_S\beta\beta$}{Interference between SM 2nu-beta-beta and Scalar-induced 2nuS-beta-beta}}

In the case of the scalar $S$ producing two active anti-neutrinos, its contribution will interfere with the standard $2\nu\beta\beta$ decay. This interference is determined by the complex-valued integral
\begin{align}
    I_\text{int}(T)  &= 
   \int \prod_{i=3}^4 \frac{d^3 \vec{p}_i}{(2\pi)^3 2E_i} 
    \frac{p_3\cdot p_4}{\left(p_3+p_4\right)^2 - m_S^2 + i \,  m_S \, \Gamma_S}  \delta\left(Q - T - E_3 - E_4\right)  \nonumber\\
    &= \frac{m_S^2 - i\,\Gamma_S m_S}{8(2\pi)^4}\int_0^{Q-T} dE_3 \Bigg\{ \frac{ 4 E_3 E_4}{m_S^2 - i\, \Gamma_S \, m_S} + \frac{1}{2} \log\left(\frac{\left(m_S^2 - 4 \,E_3 E_4  \right)^2 + \Gamma_S^2 m_S^2}{m_S^2(\Gamma_S^2 + m_S^2)}\right) \nonumber \\
    & \qquad \qquad \qquad \qquad \qquad \qquad + i \left[ \text{arccot}\left(\frac{\Gamma_S m_S}{ m_S^2 - 4 \,E_3 E_4 }\right) 
    - \text{arccot}\left(\frac{\Gamma_S}{m_S}\right)\right] 
      \Bigg\}.
\end{align}
Here, we note that the convention for $\text{arccot}(x)$ is such that it has the range $-\pi/2 < \text{arccot}(x) < \pi/2$.

The contribution of the decay rate from this interference is given by 
\begin{equation}
    \frac{d\Gamma^{2\nu}_\text{int}}{dT} = 4  \, \pi \, \kappa_{0\nu} \left(\frac{m_e}{2R}\right) 
    \text{Re}\left[a_\nu^2 \, \M_{0\nu} \, \M_{2\nu}^* \, I_\text{int}(T)\right] g(T)\, \Theta(Q - T).
\label{eq:2nbbint}
\end{equation}
It thus depends on both the $0\nu\beta\beta$ decay NME $\M_{0\nu}$ and the $2\nu\beta\beta$ decay NME $\M_{2\nu}$. As an interference between a SM and exotic contribution, it is proportional to the small exotic coupling to second power only. It is also enhanced relative to SM $2\nu\beta\beta$ decay by the appearance of the larger NME $M_{0\nu}$. We still expect that its contribution will be small as the interference requires a sizeable width $\Gamma_S$ or off-shell $S$ production, $T > Q - m_S$, which is strongly suppressed compared to the on-shell production. 


\subsection{Scalar Emission}

Taking the limit of zero width, $\Gamma_S \to 0$, and considering $S$ as a real particle in the final state, we recover the case of scalar emission $0\nu\beta\beta S$, 
\begin{equation}
    \frac{d\Gamma_S}{dT} = 
    \frac{\kappa_{0\nu}}{8\pi^2} \left(\frac{m_e}{2R}\right)^2 \left|\M_{0\nu}\right|^2 
    |a_\nu|^2 g(T) \sqrt{(Q-T)^2 - m_S^2}\,\Theta(Q - m_S - T).
\label{eq:0nbbS}
\end{equation}
We note that this limit is, strictly speaking, not self-consistent as at least the decay to massless active neutrinos is always allowed if $m_S > 0$ and thus $\Gamma_S > 0$. If all couplings involved are small, this is still a well justified approximation. For $m_S = 0$, the above expression gives the standard Majoron contribution $0\nu\beta\beta J$~\cite{Hirsch:1995in}. As we wrote our result as distributions over $T$ and $\Delta T$ instead of the electron energies, an additional factor of $1/2$ appears in the pre-factor compared to the expression commonly found in the literature~\cite{Hirsch:1995in, Brune:2018sab}.


\subsection{Standard Model \texorpdfstring{$2\nu\beta\beta$}{2nu-beta-beta} Decay}

Finally, we give the spectrum for standard $2\nu\beta\beta$ decay as the irreducible background to the exotic contributions presented above. Its calculation is non-trivial and it receives corrections to the spectral shape due to the dependence of the corresponding NME on the final lepton momenta \cite{Simkovic:2018rdz}. For consistency with the above calculation, we ignore such corrections and use the usual expression, decoupling the NME and the phase space integral,
\begin{equation}
    \frac{d\Gamma^{2\nu}_\text{SM}}{dT} = 
    \frac{\kappa_{0 \nu}}{2 \pi^2} |M_{2\nu}|^2 g(T) \frac{1}{30}\left(Q-T\right)^5
    \,\Theta(Q - T),
\label{eq:2nbb-SM}
\end{equation}
matching with the existing literature in this approximation~\cite{Simkovic:2018rdz}.


\subsection{Spectral Shapes}

\begin{figure}[t!]
    \centering
    \includegraphics[width=0.8\textwidth]{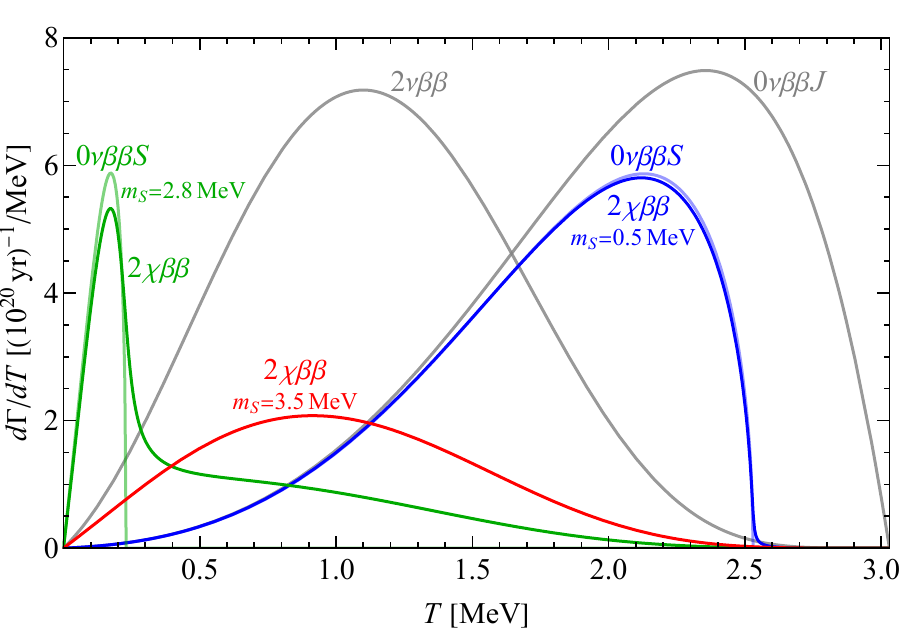}
    \caption{Differential distribution $d\Gamma/dT$ of $2\chi\beta\beta$ decay of $^{100}$Mo with respect to the total electron kinetic energy $T$ in Model I$^\prime$ for three different parameter sets: $(m_S, a_\nu, a_\chi) = (0.5~\text{MeV}, 3\times 10^{-3}, \sqrt{2})$ (blue), $(2.8~\text{MeV}, 7\times 10^{-2}, \sqrt{2})$ (green) and $(3.5~\text{MeV}, 2\times 10^{-1}, \sqrt{2})$ (red). All other couplings are set to zero and the $\chi$ are massless. For comparison, the scalar emission contributions $0\nu\beta\beta S$ are also shown, for the corresponding values of $m_S$ and $a_\nu$. The SM $2\nu\beta\beta$ and ordinary Majoron emission $0\nu\beta\beta J$ (with $a_\nu = 3\times 10^{-3}$) are shown in grey. The relevant NMEs are given in Table~\ref{tab:isotopes}.}
    \label{fig:spectra}
\end{figure}
Before analyzing the experimental sensitivity to the double beta decay contributions calculated above, we illustrate the decay spectra of the processes calculated above. In Fig.~\ref{fig:spectra}, we show the decay distribution of $2\chi\beta\beta$ for increasing values of the $s$-channel mediator mass $m_S =0.5$~MeV (blue), $2.8$~MeV (green) and $3.5$~MeV (red) in Model I$^\prime$ for massless $\chi$. The spectra, for $^{100}$Mo, having the largest $Q$ value $Q = 3.03$~MeV among the isotopes considered here, are not normalized but the couplings of $S$ to neutrinos and $\chi$ are chosen such that the features are visually enhanced: $(a_\nu, a_\chi) = (3\times 10^{-3}, \sqrt{2})$ (blue), $(7\times 10^{-2}, \sqrt{2})$ (green) and $(2\times 10^{-1}, \sqrt{2})$ (red). Thus, the coupling $a_\chi$ to $\chi$ is large and always dominant compared to $a_\nu$. The scalar $S$ therefore almost exclusively decays to $\chi$ with a branching ratio Br$(S\to\chi\chi) \approx 1$. 

This can be most easily seen by comparing with the scalar-emission spectrum $0\nu\beta\beta S$ for $m_S = 0.5$~MeV that is also shown and which almost overlaps with $2\chi\beta\beta$ as we effectively have
\begin{align}
    \frac{d\Gamma^{2\chi}_S}{dT} \approx 
    \frac{d\Gamma_S}{dT} \times \text{Br}(S\to\chi\chi),
\end{align}
in narrow-width approximation. Only a slight reduction and small tail above the scalar emission threshold $T = Q - m_S \approx 2.53$~MeV can be seen due to finite width effects from the total decay width $\Gamma_S \approx 10^{-2}$~MeV. By comparing with the ordinary (massless) Majoron emission spectrum $0\nu\beta\beta J$ (for the same $a_\nu = 3\times 10^{-3}$), we can also see the common phase space reduction due to the massive $S$.

In the green scenario, $m_S = 2.8$~MeV, $S$ is near the threshold to be produced on-shell, leading to a peak at small electron energies with a nominal threshold at $Q - m_S \approx 0.23$~MeV. In many double beta decay experiments, backgrounds become dominant in this regime and such a peak would be difficult to detect. The width of $S$ becomes sizeable, though, with $\Gamma_S \approx 0.06$~MeV and, most importantly, $2\chi\beta\beta$ decay can be induced by an off-shell $S$, i.e., for electron energies $T > Q - m_S \approx 0.23$~MeV, leading to a contribution in the higher energy region where it is more easily detectable.

Finally, in the red scenario, the scalar $S$ has a mass of $3.5$~MeV and it cannot be produced on-shell in $^{100}$Mo double beta decay. The resulting spectrum of $2\chi\beta\beta$ decay has a shape similar to SM $2\nu\beta\beta$, increasingly so with larger $m_S$ where the scalar mediation can be described by an effective $\nu\nu\chi\chi$ operator, $m_S \gg Q$~\cite{Deppisch:2020sqh}.

\begin{figure}[t!]
    \centering
    \includegraphics[width=0.8\textwidth]{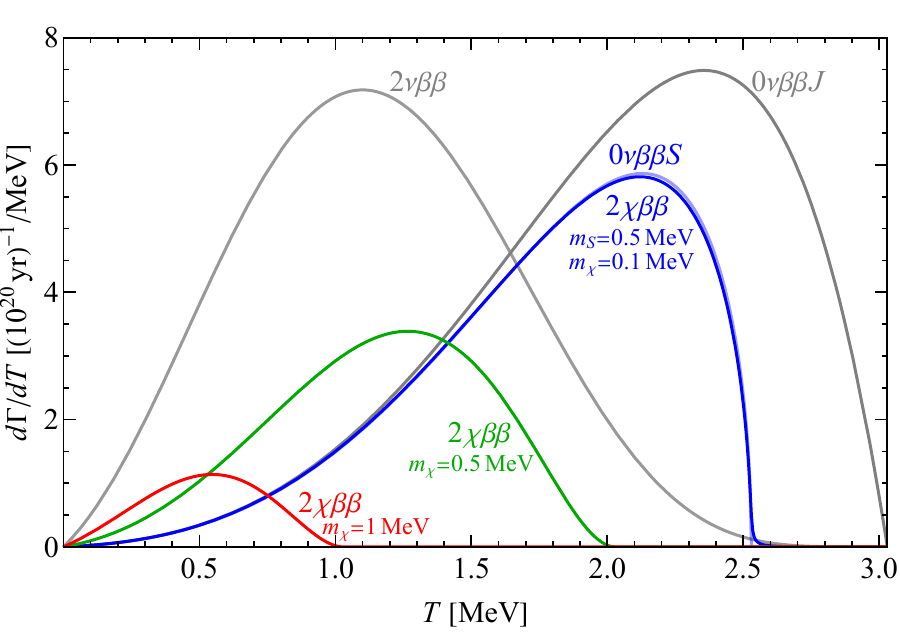}
    \caption{As Fig.~\ref{fig:spectra}, but showing $d\Gamma^{2\chi}_S/dT$ for $m_S = 0.5$~MeV and $m_\chi = 0.1$~MeV (blue), 0.5~MeV (green) and 1.0~MeV (red). The non-zero couplings are $(a_\nu, a_\chi) = (3\times 10^{-3}, \sqrt{2})$ (blue), $(7\times 10^{-2}, \sqrt{2})$ (green) and $(3\times 10^{-1}, \sqrt{2})$ (red).}
    \label{fig:spectra-massivechi}
\end{figure}
In Fig.~\ref{fig:spectra}, we always had $m_\chi = 0$. Instead, in Fig.~\ref{fig:spectra-massivechi} we show the dependence of the $2\chi_S\beta\beta$ spectrum on $m_\chi$ for a fixed $m_S = 0.5$~MeV. The blue, green and red distributions are for $m_\chi = 0.1$~MeV, 0.5~MeV and 1.0~MeV, respectively, where the $a_\nu$ coupling is changed accordingly to show a visible spectrum. In the blue case, $S$ can still be produced on-shell and there is no discernible effect compared to the massless $\chi$ case in Fig~\ref{fig:spectra}. On the other hand, for $m_\chi = 0.5$~MeV and 1.0~MeV, the mass threshold for $\chi$ production is more restrictive than that for $S$. The scalar can thus not be produced on-shell. Because of this, the depicted spectra are expected to be difficult to observe as they compete with the always accessible, and resonantly enhanced, $2\nu_S\beta\beta$ contribution.

\begin{figure}[t!]
    \centering
    \includegraphics[width=0.8\textwidth]{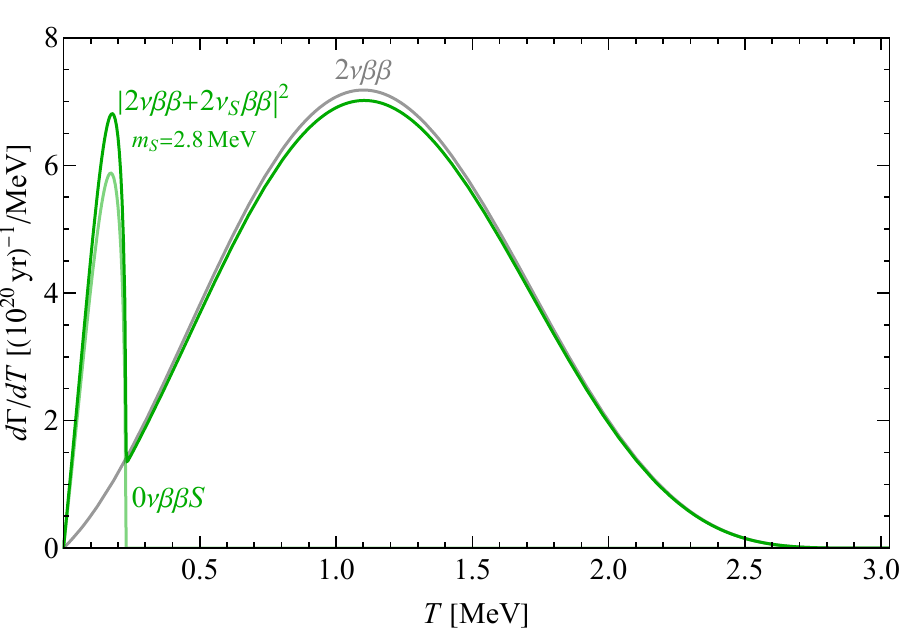}
    \caption{Differential distribution $d\Gamma/dT$ of $2\nu\beta\beta$ decay of $^{100}$Mo with respect to the total electron kinetic energy $T$ including contributions from SM $2\nu\beta\beta$, $S$-mediated $2\nu_S\beta\beta$ and their interference (green). The scalar mass is $m_S = 2.8$~MeV and $a_\nu = 7\times 10^{-2}$ with all other couplings set to zero. For comparison, the scalar emission contribution $0\nu\beta\beta S$ (using the same parameters) and SM $2\nu\beta\beta$ are also shown. The relevant NMEs are given in Table~\ref{tab:isotopes}.}
    \label{fig:spectra-nu}
\end{figure}
In Fig.~\ref{fig:spectra-nu}, we show an example of the full $2\nu\beta\beta$ decay spectrum, including contributions from SM $2\nu\beta\beta$, $S$-mediated $2\nu_S\beta\beta$ and their interference (green). The scalar mass is $m_S = 2.8$~MeV and the coupling is $a_\nu = 7\times 10^{-2}$. All other couplings are zero, i.e., there is no coupling to $\chi$, and $S$ exclusively decays to neutrinos in this case. The contributions add as
\begin{align}
    \frac{d\Gamma^{2\nu}}{dT} &=
      |\mathcal{M}_{2\nu}|^2\frac{d\tilde\Gamma^{2\nu}_\text{SM}}{dT} \nonumber\\
    &+ |a_\nu|^2 |\mathcal{M}_{0\nu}|^2 \frac{d\tilde\Gamma^{2\nu}_{S~\text{on-shell}}}{dT}
    + |a_\nu|^4 |\mathcal{M}_{0\nu}|^2 \frac{d\tilde\Gamma^{2\nu}_{S~\text{off-shell}}}{dT} \nonumber\\
    &+ 2\,\text{Re}\left[a_\nu^2 \mathcal{M}_{0\nu} \mathcal{M}_{2\nu}^*\right] \frac{d\tilde\Gamma^{2\nu}_\text{int}}{dT},
\end{align}
where we have extracted the dependence on $a_\nu$ and the NMEs with the remaining parts in the reduced width denoted by a tilde. We have also split up the on- and off-shell parts of the $S$-mediated contributions, where the former scales with $\Gamma_S^{-1}$, leading to an overall proportionality with $|a_\nu|^2$ in this case. We see that the interference depends on both the $0\nu\beta\beta$ and $2\nu\beta\beta$ NMEs and their relative signs. We here take both NMEs to be positive which leads to a negative interference above the threshold and a suppression of the spectrum compared to pure SM $2\nu\beta\beta$ decay.


\section{Cosmology and Laboratory Constraints}
\label{sec:DM}

Interactions between neutrinos and other light particles produce a range of observable effects which can be constrained by imposing astrophysical, cosmological and laboratory bounds. We begin by reviewing the existing constraints on the frameworks presented in Sec.~\ref{sec:UV}, specifically Model~I$^{\prime}$ as the exotic Dirac fermion $\chi$ is stable under the presence of a discrete $Z_2$ symmetry. These include limits on neutrino and exotic fermion (self-)interactions as well as their couplings to mediator particles. The constraints arise from a variety of sources, including cosmological observations such as those from Cosmic Microwave Background (CMB), Big Bang Nucleosynthesis (BBN), collisional damping, and thermal relic density considerations, as well as laboratory searches involving meson decays.


\subsection{Cosmology Constraints}

\subsubsection{Thermal Relic}
Considering equal number densities for DM particles and their antiparticles, the thermally averaged annihilation cross section (multiplied by the relative velocity) required to account for the observed DM abundance in the present day universe, which can be expressed by a series expansion in relative velocity $v_r$ (considering non-relativistic DM scenario with $v_r \ll c$) is \cite{Wells:1994qy, Olivares-DelCampo:2017feq}
\begin{align}
    \langle \sigma v_r \rangle = 
    a_r + b_r v_{r}^2 + d_r v_{r}^4.
\end{align}
To avoid DM overabundance, the upper limit on the DM thermal relic density \cite{Planck:2018vyg}, $\Omega_\chi h^2 \leq 0.120 \pm 0.001$ can be approximately translated to a lower limit \cite{Kolb:1990vq, Scherrer:1985zt} on $\langle \sigma v_r \rangle \geq 3\times 10^{-26} \rm{cm^3}/\rm{s}$ for a constant cross section at leading order in the expansion stated above, where $\sigma$ corresponds to the annihilation cross section of DM species to active neutrino pair, $\chi\chi \to\nu\nu$. Additionally, if the cross section, at leading-order, depends on $v_r^2$ or $v_r^4$, the required value at freeze-out will be $\langle \sigma v_r \rangle \simeq 6 \times  10^{-29}\ \rm{cm^3}/\rm{s}$ or $\langle \sigma v_r \rangle \simeq 9 \times 10^{-26} \ \rm{cm^3}/\rm{s}$, respectively~\cite{Giacchino:2013bta}.

\subsubsection{Collisional Damping} 
Interactions between neutrinos and DM can significantly affect the distribution of matter and radiation in the Universe, particularly by impacting the CMB spectrum and the formation of large-scale structures (LSS) \cite{Boehm:2000gq, Boehm:2001hm, Boehm:2004th}. There are two primary effects to consider: DM is no longer collisionless due to its interaction with neutrinos and the mediator particle, and the neutrinos remain collisional for a longer period compared to the standard cosmological scenario. This reduces the length of DM free streaming and enhances their ability to cluster more efficiently which eventually contributes to the formation of large-scale structures. By comparing theoretical predictions with CMB and LSS observations, constraints can be placed on the strength of $\nu$-DM interactions. In this regard, Planck data suggests that the $\nu$-DM elastic scattering cross section for the process $\nu\chi\to\nu\chi$ must be \cite{Akita:2023yga,Dev:2025tdv}
\begin{equation}
    \sigma_{\text{el}} < 
    10^{-29} \left (\frac{m_\chi}{\rm{MeV}}\right)~ \text{cm}^2.
\end{equation}

\subsubsection{Neutrino Reheating Bounds} 
DM annihilation into neutrinos after they have decoupled from electrons, at temperatures $T \lesssim T_{D} \sim 1$ MeV (where $T_D$ is the decoupling temperature of DM from electrons) \cite{Kolb:1986nf, Serpico:2004nm, Boehm:2012gr, Ho:2012ug, Steigman:2013yua, Boehm:2013jpa, Nollett:2013pwa, Steigman:2014pfa, Nollett:2014lwa}, can reheat the neutrino sector, producing observable effects. This change in neutrino energy density effectively increases the number of relativistic degrees of freedom, $N_{\rm{eff}}$, in the early Universe which can be expressed as
\begin{equation}
    \frac{\rho_\nu}{\rho_\gamma} \equiv 
    1 + \frac{7}{8} \left(\frac{T_\nu}{T_\gamma}\right)^{4/3} N_\text{eff},
\end{equation}
where $\rho_\gamma$ is the photon energy density and $T_\nu, T_\gamma$ correspond to the temperature of neutrinos and photons, respectively. In our framework, because of interactions with neutrinos, the DM particle remains in thermal equilibrium for a longer time. Therefore, the effective number of light relativistic species becomes \cite{Boehm:2013jpa, Wilkinson:2016gsy}
\begin{align}
    N_\text{eff} = 
    N_\nu \left[1 + \frac{1}{N_\nu} \frac{g_\chi}{2} F\left(\frac{m_\chi}{T_D}\right)^{4/3} \right],
\end{align}
where $N_\nu = 3.046$ \cite{Mangano:2001iu, Mangano:2005cc} and $g_\chi = 4$ as the considered DM is a Dirac particle. The function $F(x)$ is defined as
\begin{align}
    F(x) = \frac{30}{7\pi^4} 
    \int_x^\infty dy \frac{(4y^2 - x^2)\sqrt{y^2 - x^2}}{e^y \pm 1}.
\end{align}

\subsubsection{CMB-NSI} 
Neutrino non-standard interactions (NSI) influence the epoch of neutrino free streaming and their coupling to the photon–baryon fluid in the early Universe. Such effects manifest as modifications in the phase shift and amplitude of the matter power spectrum, leading to stringent cosmological bounds. These constraints apply not only to neutrino self-interactions but also to interactions with a light dark sector, with implications for alleviating long-standing tensions in cosmological data, such as the $H_0$ and $\sum m_\nu$ tensions \cite{Dev:2025tdv}. The most recent CMB analysis excludes effective four-neutrino interactions $G_\text{eff}(\bar\nu \nu)(\bar\nu \nu)$ as \cite{Camarena:2024daj}
\begin{equation}
    G_{\text{eff}} \simeq 
    \frac{y_\nu^2}{m_{S}^2} < 5.6 \times 10^{-5}~\text{MeV}^{-2},
\end{equation}
where $y_\nu$ corresponds to a generic coupling of neutrinos with a mediator of mass $m_S$.

\subsubsection{CMB-\texorpdfstring{$\nu$}{nu}DM}
Additional constraints on $\nu$-DM interactions arise from cosmological observables including the Lyman-$\alpha$ forest \cite{Wilkinson:2014ksa}, structure formation \cite{Crumrine:2024sdn} and small-scale CMB data \cite{Giare:2023qqn}. The upper limits on the DM-neutrino scattering cross section can be parameterized as \cite{Wilkinson:2014ksa}
\begin{equation}
    \sigma_{\nu\chi} < \sigma_n 
    \left( \frac{m_\chi}{\text{GeV}} \right) 
    \left( \frac{T_\nu}{T^0_\nu} \right)^n,
\end{equation}
with $T^0_\nu = 6.1~\text{K}$, $\sigma_{0} = 2\times 10^{-28}~\text{cm}^2$, $\sigma_2 = 2\times 10^{-39}~\text{cm}^2$ and $n=2$ \cite{Dev:2025tdv}.

\subsubsection{BBN} 
If both the mediator and light particles are at the MeV scale, additional constraints arise from matching with the observed Big Bang Nucleosynthesis (BBN) parameters. The Hubble rate at the time of BBN, affected by the presence of neutrinos and other light degrees of freedom modifies the neutron-to-proton ratio and the freeze-out of
deuterium reactions. This constrains the number of additional effective light degrees of freedom around the time of BBN and neutrino decoupling $T \approx 1-2$~MeV. The determination of a limit on the scalar mass $m_S$ is non-trivial. It is further modified by the fact that neutrinos self-interact via $S$. In Ref.~\cite{Blinov:2019gcj} the weakest limit was reported in the moderately interacting self-interacting neutrino scenario, with $a_\nu^2 / m_S^2 \approx (100~\text{MeV})^{-2}$, which is near the sensitivity of double beta decay searches reported below. In this case, the upper limit is $m_S < 0.6$~MeV and 3.7~MeV at 95\% confidence level, for a real and complex scalar $S$, respectively. As $S$ is charged in our case, the more restrictive limit should apply but we note that due to the neutrino-number violating nature of our interactions, $S$ may effectively act as a real scalar, if there is a sizeable neutrino asymmetry present.

The above constraint applies as long as $S$ is in thermal equilibration with the SM plasma at $T \approx 1-2$~MeV via $\nu\nu \leftrightarrow S$. This is the case if $\Gamma(\nu\nu \leftrightarrow S) / H > 1$ with the reaction rate $\Gamma(\nu\nu \leftrightarrow S) = g K_1(m_S/T) / K_2(m_S / T) a_\nu^2 m_S / (32\pi)$ and the Hubble expansion rate $H = 1.66\sqrt{g_*} T^2/m_\text{Planck}$ where $g=1$ (2) for a real (complex) scalar, $K_i(z)$ are Bessel functions, $g_* = 10.75$ and we take the temperature $T = \text{max}(1.5~\text{MeV}, m_S)$. This is satisfied for most of the parameter space considered here, i.e., $S$ can be considered in thermal equilibrium at the time of BBN. Only for small couplings and masses, $g (a_\nu / 8\times 10^{-7})^2 (m_S / \text{keV})^2 \lesssim 1$, this is not the case, and $S$ will not be produced abundantly.

In our scenario with a coupling to a dark fermion $\chi$, the above considerations are further complicated, as it will modify the injection of neutrinos at the time of BBN. An exact treatment is beyond the scope of this work. To indicate the relevant parameter space where this effect should become important, we consider the process $\nu\nu\to\chi\chi$, via $s$-channel scalar mediation, that keeps the dark fermion in equilibrium with the SM plasma. The interaction rate is given by $\Gamma_{\nu\chi} = \sigma(\nu\nu\to S\to\chi\chi) n_\nu(T)$, where $n_\nu(T)$ is the neutrino number density at temperature $T$. We have considered that the neutrinos remain in thermal equilibrium with the SM plasma around this temperature. The process is effective if $\Gamma_{\nu\chi} / H < 1$ \cite{Dev:2025tdv} which we will indicate below.


\subsection{Constraints from Supernovae}

The coupling of the light scalar $S$ with neutrinos can further be constrained considering the significant impact of this coupling on the mechanism and cooling of supernovae (SN). We mostly focus on two different constraints from SN data. Firstly, the binding energy released in a SN is well predicted with the emitted neutrino signal measured from SN1987A \cite{Kamiokande-II:1987idp, Hirata:1988ad, Alekseev:1988gp, Alekseev:1987ej, Bionta:1987qt}. The presence of a light scalar $S$ and its coupling with active neutrinos will affect the emitted luminosity, but this must not alter the experimental observation. This puts a stringent constraint on such a coupling \cite{Brune:2018sab}. Secondly, if $S$ couples too strongly it may be trapped within the SN core \cite{Brune:2018sab}. However, the scalar $S$ couples to a dark fermion $\chi$ in our scenario, potentially strongly, where such bounds are expected to be weakened. Therefore, we will not consider this trapping constraint further.


\subsection{Meson Decays}

For a light scalar mediator $S$ coupled to neutrinos and an exotic fermion, rare meson decays such as kaon decays, $K^\pm \to \ell^\pm + \nu/\chi + S$ provide strong constraints on the relevant couplings involved \cite{Pasquini:2015fjv}. At tree level, these three-body decays suffer from infrared divergences in the limit $m_S \to 0$, with the divergent terms proportional to $m_\ell^2 \log(m_S^2/m_K^2)$. The divergences cancel one-loop corrections to the standard 2-body decay $K^\pm \to \ell^\pm \nu$, in accordance with the Kinoshita-Lee-Nauenberg theorem. Using kaon partial width data \cite{ParticleDataGroup:2024cfk}, stringent bounds can be placed on scalar-neutrino or scalar-exotic fermion interactions \cite{Berryman:2018ogk,  Dev:2025tdv}. We specifically consider the rare Kaon decay to constrain the parameter space as it gives the most stringent limit on the couplings for our masses of interest. There are other laboratory constraints, not competitive in our parameter space considered, arising from invisible $Z$ decays, modification of the Fermi constant $G_F$ and others \cite{Dev:2025tdv}.


\subsection{Impact on Parameter Space}

With the above considerations, we summarize the impact on the parameter space. For Model I$^\prime$ in our study, the relevant cosmological processes are DM self-annihilation to neutrino pair $\chi\chi\to\nu\nu$, DM-neutrino scattering $\nu \chi\to\nu\chi$ as well as $S$ and $\chi$ production from neutrinos $\nu\nu\to S$,  $\nu\nu\to\chi\chi$. The relevant processes are discussed above and matrix elements are given in Appendix~\ref{sec:appendix}.

\begin{figure}[t!]
    \centering
    \includegraphics[width=0.7\textwidth]{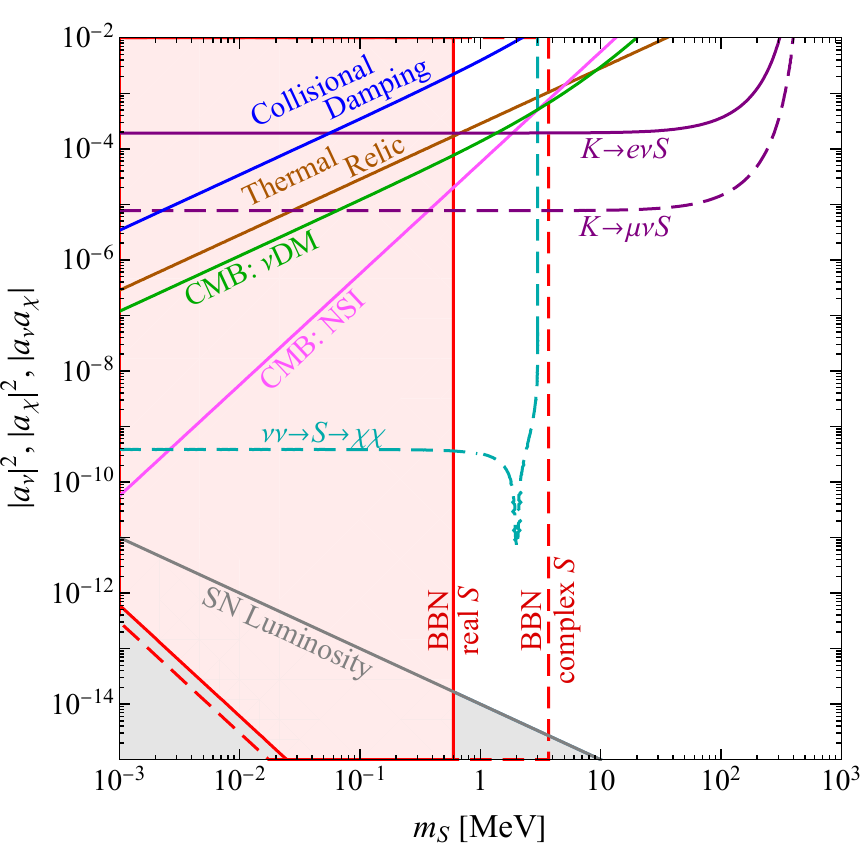}
    \caption{Constraints on (products of) couplings from cosmological (DM thermal relic, collisional damping, CMB and BBN), supernovae and kaon decay bounds as a function of the scalar mass $m_S$. The dark fermion $\chi$ mass is set at $m_\chi = m_S/3$.}
    \label{fig:cosmo_lab}
\end{figure}
Fig.~\ref{fig:cosmo_lab} shows constraints on couplings, or products of couplings, as a function of the scalar mass $m_S$. The combinations of couplings $|a_\nu|^2, |a_\chi|^2, |a_\nu a_\chi|$ are involved in different processes, in the simplified parameter space where we set $b_{\nu,\chi} = 0$. We also set $m_\chi = m_S/3$. The curve labeled ``Thermal Relic" shows the coupling required to achieve the observed relic density for $\chi$. Cosmological limits arise from the CMB, with ``CMB: $\nu$DM" and ``CMB: NSI" curves indicating upper limits where scalar interactions distort the CMB anisotropy spectrum. Likewise, the curve labelled ``Collisional Damping" corresponds to excessive interactions of $S$ with neutrinos and DM $\chi$ interrupts structure formation. The most restrictive laboratory constraints in the parameter space shown, apart from double beta decay searches, are from rare kaon decays, $K\to e\nu S$ and $K\to\mu\nu S$, which probe couplings down to $\sim 10^{-2}$ and $\sim 10^{-2.5}$, respectively, for $m_S \lesssim 100~\text{MeV}$. We note that $K\to \mu\nu S$ involves coupling to $\nu_\mu$, i.e., not directly comparable to the electron coupling involved in double beta decay. The red region labelled ``BBN, real $S$" is excluded by constraints on the presence of a light real scalar, if it has thermalized by the time of BBN. The dashed-line bounded region ``BBN, complex $S$" gives the corresponding constraint for a complex scalar. These BBN constraints were determined assuming the presence of only an additional scalar coupling with neutrinos, i.e., without a dark fermion $\chi$. They are expected to be modified if $\chi$ thermalizes as well, which occurs above the curve $\nu\nu\to S\to\chi\chi$ in the described scenario as an example. Finally, we also show the constraint from SN luminosity considerations \cite{Brune:2018sab, Telalovic:2024cot} where the grey shaded region at the bottom is excluded as a result of recasting the limits derived in \cite{Telalovic:2024cot} in our scenario.


\section{Statistical Procedure}
\label{sec:statistics}

Our goal is to estimate the sensitivity of current and future double beta decay experiments to the different models discussed in Sec.~\ref{sec:UV}, and we follow the standard frequentist approach~\cite{Tanabashi:2018oca}. The total double beta decay rate with respect to the electrons' kinetic energy~$T$ in our case is then generally written as
\begin{align}
\label{eq:total_dist}
	\frac{d\Gamma(\mathbf x)}{dT} 
	= \frac{d\Gamma^{2\nu}_\text{SM}}{dT} 
	+ \frac{d\Gamma_\text{BSM}(\bf x)}{dT}.
\end{align}
Here, $d\Gamma^{2\nu}_\text{SM}/dT$ represents the standard $2\nu\beta\beta$ decay contribution in the SM, which will act as background to the search for an exotic contribution $d\Gamma_\text{BSM}(\mathbf x) / dT$, where $\mathbf x$ represents the parameters of the model in question. For example, in the simplest scenario of scalar emission, we have $\mathbf x = (a_\nu, m_S)$. The BSM contribution will vanish in a certain parameter limit $\mathbf x \to \mathbf 0$, e.g., for $a_\nu \to 0$ in the scalar emission case.

Experiments observe events of double electron emission distributed over energy bins. Theoretically, the expected number of events within an energy interval $T_i < T < T_i + \Delta T$ is calculated as
\begin{align}
	N^{(i)}_\text{exp} (\mathbf x)
	= N_A m_\text{iso}^{-1} \mathcal{E} 
    \int_{T_i}^{T_i + \Delta T} dT \frac{d\Gamma(\mathbf x)}{dT}. 
\end{align}
The factor in front of the integral gives the effective number of decaying nuclei with Avogadro's number $N_A = 6.022\times 10^{23}$~mol$^{-1}$, the molar mass $m_\text{iso}$ of the isotope in question, and the fiducial exposure $\mathcal{E}$ of the experiment in units of kg$\cdot$yr.

Having observed a set of events, the log-likelihood, or $\chi^2(\mathbf x)$, of the binned data $\mathbf D = \{N_\text{obs}^{(i)}\}$ given a BSM hypothesis is
\begin{align}
\label{eq:log_likelihood}
	-2\log\mathcal{L}(\mathbf D|\mathbf x) 
	&= 2\sum^{N_\text{bins}}_i
	   \left\{N^{(i)}_\text{exp}(\mathbf x) - N^{(i)}_\text{obs} 
	        + N^{(i)}_\text{obs}
	          \log\left(
	          \frac{N^{(i)}_\text{obs}}{N^{(i)}_\text{exp}(\mathbf x)}
            \right)\right\} 
	\nonumber\\
	&\approx \chi^2(\mathbf x) = \sum^{N_\text{bins}}_i 
	\frac{\left(
        N^{(i)}_\text{obs} - 
	   N^{(i)}_\text{exp}(\mathbf x)\right)^2}{N^{(i)}_\text{exp}(\mathbf x)},
\end{align}
where the second line applies if there are sufficiently many events per bin \cite{Wilks:1938dza}. For the standard $2\nu\beta\beta$ decay as common contribution, this is the case for the current and future experimental exposures considered below.

In order to estimate the sensitivity of an experiment, we assume that it does not see a distribution that differs significantly from the SM prediction, setting the number of observed events to $N^{(i)}_\text{obs} = N^{(i)}_\text{exp}(\mathbf 0)$. This represents the so-called Asimov data set \cite{Burns:2011xf} which provides a good approximation to Monte Carlo simulations of the experiment that would otherwise be required \cite{Cowan:2010js}.

In addition to the statistical uncertainty, we include two types of systematic errors:
\begin{enumerate}
    \item Systematic experimental errors, assumed proportional to the number of events in a bin, $\sigma_\text{sys}^{(i)} = \sigma_f N_\text{exp}^{(i)}$, are added in quadrature to the statistical uncertainty. We take $\sigma_f$ as constant for a given experiment.

    \item The main theoretical uncertainty arises from the $2\nu\beta\beta$ and $0\nu\beta\beta$ NMEs that enter the decay rate calculations. The standard $2\nu\beta\beta$ decay depends on $\mathcal{M}_{2\nu}$ whereas the BSM contributions, due to their ordinary Majoron structure, mainly scale with $\mathcal{M}_{0\nu}$ (the interference of the scalar-mediated $2\nu_S\beta\beta$ with SM $2\nu\beta\beta$ will also depend on $\mathcal{M}_{2\nu}$). Calculations in different nuclear structure models differ by factors of $2-3$ in a given isotope. In order to include this effect, we treat both $\mathcal{M}_{2\nu}$ and $\mathcal{M}_{0\nu}$ as Gaussian-distributed nuisance parameters with best fit values and standard deviations, $\langle \mathcal{M}_{n\nu}\rangle \pm \delta \mathcal{M}_{n\nu}$, derived from a survey of NME values \cite{Agostini:2022zub}. We also include a potential correlation between $\mathcal{M}_{2\nu}$ and $\mathcal{M}_{0\nu}$. While this correlation is not direct and perfect, it is observed statistically in several NME calculations and is encoded by a statistical correlation factor $\rho$ \cite{Jokiniemi:2022ayc}. Jointly, $\mathcal{M}_{2\nu}$ and $\mathcal{M}_{0\nu}$ follow a bi-variate, potentially tilted, Gaussian distribution.
\end{enumerate}

We thus have the $\chi^2$ function
\begin{align}
	\chi^2(\mathbf x) &=
    \sum^{N_\text{bins}}_i 
	\frac{\left(
          N^{(i)}_\text{exp}(\mathbf 0) 
        - N^{(i)}_\text{exp}(\mathbf x)
    \right)^2}{N^{(i)}_\text{exp}(\mathbf x) + (\sigma_\text{sys}^{(i)})^2} 
    \nonumber\\
    &+ \frac{1}{1-\rho^2}
    \left[
          \frac{(\eta - \langle\mathcal{M}_{2\nu}\rangle)^2}{(\delta \mathcal{M}_{2\nu})^2}
        + \frac{(\kappa - \langle \mathcal{M}_{0\nu}\rangle)^2}{(\delta \mathcal{M}_{0\nu})^2}
        - 2\rho\frac{(\eta - \langle\mathcal{M}_{2\nu}\rangle)(\kappa - \langle\mathcal{M}_{0\nu}\rangle)}{\delta \mathcal{M}_{2\nu} \cdot \delta \mathcal{M}_{0\nu}}
    \right],
\end{align}
where the set of parameters $\mathbf x$ now includes the two NME nuisance parameters $\eta$, $\kappa$, and the Asimov data set is calculated using the best fit value $\eta = \langle \mathcal{M}_{2\nu}\rangle$. By construction, $\chi^2(\mathbf 0) = 0$ for the Asimov data set, i.e., the SM $2\nu\beta\beta$ decay case with $\langle \mathcal{M}_{2\nu}\rangle$ as NME.

We are mainly interested in the general sensitivity to the BSM scenarios considered and we do not attempt to fit all model parameters at the same time. Instead, we choose one coupling constant $g$ as fitting parameter while keeping all other parameters fixed, specifically the particle masses involved. For example, in the simple scalar emission model, we calculate $\chi^2$ using $a_\nu$ for a fixed scalar mass $m_S$. In order to find the 90\% confidence level (CL) limit on such a coupling constant $g$, we first minimize over the NME nuisance parameters,
\begin{align}
    \chi^2(g) = 
    \underset{\eta, \kappa}{\text{min}}\left[\chi^2(g, \eta, \kappa)\right].
\end{align}
This profile function then follows a chi-squared distribution with one degree of freedom and to find the 90\% CL upper limit on $g$, we set $\chi^2(g) = 2.71$.


\section{Sensitivity of Double Beta Decay Experiments}
\label{sec:results}

\begin{table}[t!]
	\setlength{\tabcolsep}{6pt}
	\begin{tabular}{cccc|cccc}
		\hline
		Isotope & $m_\text{iso}$~[g/mol] & $Q$~[MeV] & $(T_{1/2}^{2\nu})_\text{exp}$~[yr] & $\langle \mathcal{M}_{2\nu}\rangle$ & $\delta\mathcal{M}_{2\nu}$ & $\langle \mathcal{M}_{0\nu}\rangle$ & $\delta\mathcal{M}_{0\nu}$ \\\hline
		$^{76}_{32}$Ge & 72.6 & 2.04 & $1.9\times 10^{21}$ &0.10 & 0.22 & 4.3 &1.2 \\
		$^{100}_{\phantom{1}42}$Mo & 96.0 & 3.03 & $7.1\times 10^{18}$ & 0.19 & 0.27 & 4.9 & 1.5 \\
		$^{136}_{\phantom{1}54}$Xe & 131 & 2.46 & $2.2\times 10^{21}$ & 0.016 & 0.19 & 2.6 & 1.1 \\\hline
	\end{tabular}
	\caption{Molar mass $m_\text{iso}$, $Q$ value and experimental $2\nu\beta\beta$ decay half-life $(T_{1/2}^{2\nu})_\text{exp}$ for the isotopes considered in this work. Also given are the average NMEs $\langle \mathcal{M}_{2\nu}\rangle$, $\langle \mathcal{M}_{0\nu}\rangle$ and the standard deviations $\delta\mathcal{M}_{2\nu}$, $\delta\mathcal{M}_{0\nu}$.}
	\label{tab:isotopes}
\end{table}
In Tab.~\ref{tab:isotopes}, we provide the relevant data for the three isotopes considered in this work, $^{76}$Ge, $^{100}$Mo and $^{136}$Xe. As described above, we treat the $2\nu\beta\beta$ and $0\nu\beta\beta$ NMEs as individual random variables, based on the spread in different nuclear structure calculations. We fix the best fit value $\langle \mathcal{M}_{2\nu}\rangle$ using the experimentally measured $2\nu\beta\beta$ decay half lives via Eq.~\eqref{eq:2nbb-SM}. For the NME $\mathcal{M}_{0\nu}$, we compute the mean of the values provided in Table~I of \cite{Agostini:2022zub} which includes calculations from the Shell Model, QRPA, EDF and IBM frameworks. The corresponding uncertainty $\delta\mathcal{M}_{0\nu}$ is evaluated as the standard deviation of this set, providing a measure of the spread among theoretical predictions. As discussed in \cite{Jokiniemi:2022ayc}, the matrix elements $\mathcal{M}_{0\nu}$ and $\mathcal{M}_{2\nu}$ are potentially correlated and can be approximated by a linear relation, $\mathcal{M}_{2\nu}/q^2 = a + b\cdot  A^{-1/6}\mathcal{M}_{0\nu}$, where $a$ and $b$ are real constants and $q$ is a factor capturing the potential axial-coupling quenching. We incorporate this correlation into our uncertainty estimates, by propagating the variation in $\mathcal{M}_{0\nu}$ through this expression. The resulting uncertainty in $\mathcal{M}_{2\nu}$ is therefore given by
\begin{equation}
    \delta\mathcal{M}_{2\nu} = b A^{-1/6}\cdot \delta\mathcal{M}_{0\nu},
\end{equation}
where we take a typical value $b = 0.18$ for the slope of the proportionality \cite{Jokiniemi:2022ayc}.

In the following, we will assume two different NME scenarios:
\begin{enumerate}
    \item For the existing experimental searches at GERDA II, NEMO-3 and KamLAND-Zen, we assume the NME values in Tab.~\ref{tab:isotopes}, without correlation between $\mathcal{M}_{2\nu}$ and $\mathcal{M}_{0\nu}$, $\rho = 0$.

    \item For the future searches at LEGEND-1000, CUPID and nEXO, we take the same values above as a pessimistic scenario (no future improvement in NME calculations) but also consider an optimistic scenario as follows: We keep the best fit values from Tab.~\ref{tab:isotopes} but reduce the uncertainties by a factor of five, $\delta\mathcal{M}_{n\nu} \to 0.2\times \delta\mathcal{M}_{n\nu}$. At the same time, we include a statistical correlation between $\mathcal{M}_{0\nu}$ and $\mathcal{M}_{2\nu}$, with a correlation factor $\rho = 0.85$ \cite{Jokiniemi:2022ayc}. This is intended as an optimistic choice for the state of NME calculations at the time of the experimental search. 
\end{enumerate}

\begin{table}[t!]
	\setlength{\tabcolsep}{6pt}
	\begin{tabular}{cccccc}
		\hline
		Isotope & Experiment & $\mathcal{E}$~[$\text{kg}\cdot\text{yr}$] & $T_\text{min}$~[keV] &
        $\Delta T$~[keV] & $\sigma_f$~[\%] \\\hline
		\multirow{2}{*}{$^{76}$Ge}& GERDA II \cite{GERDA:2020xhi} & $104$ & 500 &
        $15$ & $1.9$  \\
		& LEGEND-1000 \cite{Zsigmond:2020bfx} & $7.0\times 10^3$ & 100 &
        $2.5$ & $0.5$ \\\hline
		\multirow{2}{*}{$^{100}$Mo}& NEMO-3 \cite{NEMO-3:2019gwo} & $34$ & 500 & 
        $100$ & $1.8$ \\
		& CUPID-1T \cite{CUPIDInterestGroup:2019inu} & $1.7\times 10^3$ & 100 &
        $5$ & $0.5$ \\\hline
		\multirow{2}{*}{$^{136}$Xe}& KamLAND-Zen \cite{KamLAND-Zen:2019imh} & $126$ & 100 & $50$ & $0.3$ \\
		& nEXO \cite{nEXO:2021ujk} & $1.9\times 10^4$ & 100 & $5$ & $0.5$ \\\hline
	\end{tabular}
	\caption{Current or expected future exposure $\mathcal{E}$, minimum electron energy $T_\text{min}$, energy resolution (bin width) $\Delta T$ and relative systematic uncertainty $\sigma_f$ for the double beta decay experiments considered in this work.}
	\label{tab:experiments}
\end{table}
We will estimate the sensitivity for different experimental setups as described in Tab.~\ref{tab:experiments}. For each of the relevant isotopes $^{76}$Ge, $^{100}$Mo and $^{136}$Xe, we consider an existing $0\nu\beta\beta$ decay search (at GERDA~II, NEMO-3 and KamLAND-Zen, respectively) and at a future planned experiment (LEGEND-1000, CUPID and nEXO, respectively). We have extracted the key experimental metrics for our statistical approach from the references listed in Table~\ref{tab:experiments} and \cite{Agostini:2022zub}, specifically, the assumed exposure $\mathcal{E}$, the minimum usable electron energy $T_\text{min}$, the energy resolution (bin width) $\Delta T$ and the relative systematic uncertainty $\sigma_f$. In determining the binned spectrum, we also exclude the highest $100$~keV below the $2\nu\beta\beta$ endpoint, mainly to ensure that there are enough events in each bin to justify the use of the $\chi^2$ function in Eq.~\eqref{eq:log_likelihood}. We therefore assume a usable spectrum over the interval $T_\text{min} < T < Q - 0.1$~MeV. Our statistical approach described in Sec.~\ref{sec:statistics} is not intended to exactly reproduce the experimental sensitivities, but to provide an estimate that mainly incorporates the statistical uncertainty in measuring the SM $2\nu\beta\beta$ spectrum, with some theoretical and experimental systematic errors captured.


\subsection{Scalar Emission}

We first consider simple scalar particle emission, $0\nu\beta\beta S$ as the BSM signal, i.e., $S$ is assumed to be emitted as a real particle in the final state. The measured electron energy spectrum is then given by
\begin{align}
    \frac{d\Gamma(\mathbf{x})}{dT} = 
    \frac{d\Gamma^{2\nu}_\text{SM}}{dT}
    + \frac{d\Gamma_S(\mathbf{x})}{dT},
\end{align}
with the SM $2\nu\beta\beta$ and $0\nu\beta\beta S$ spectra given in Eqs.~\eqref{eq:2nbb-SM} and Eq.~\eqref{eq:0nbbS}, respectively. The model parameters are the scalar mass $m_S$ and the coupling $a_\nu$ to active neutrinos, $\mathbf{x} = (m_S, a_\nu)$.

\begin{figure}[t!]
    \centering
    \includegraphics[width=0.8\textwidth]{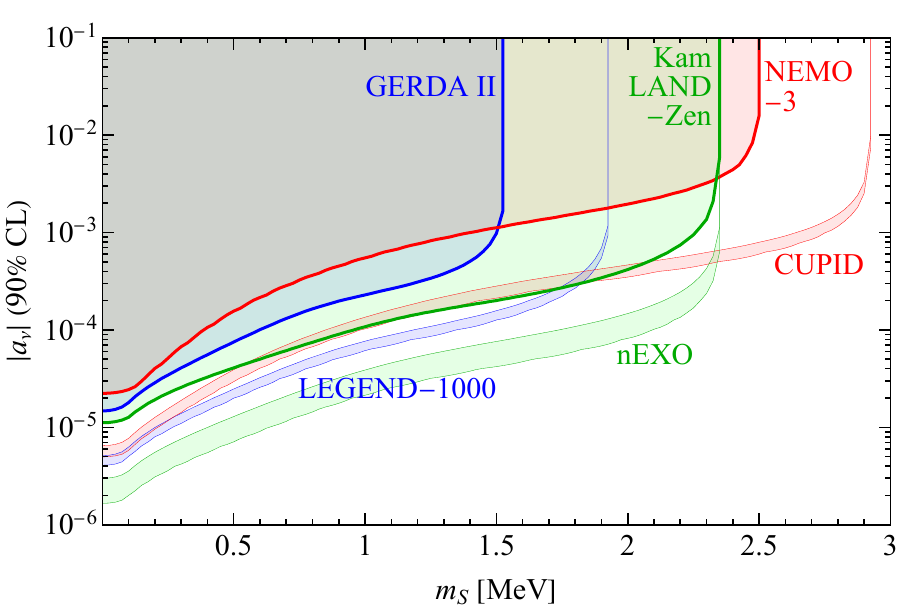}
    \caption{Estimated upper limits and sensitivity on the coupling $|a_\nu|$ at 90\%~CL in scalar emission $0\nu\beta\beta S$ decay as a function of the scalar mass $m_S$. The experimental setups for the current and planned future experiments are given in Table~\ref{tab:experiments}. For the sensitivity of future experiments, the coloured bands indicate the use of two sets of NMEs: A pessimistic scenario with current, uncorrelated uncertainties as given in Table~\ref{tab:isotopes}; and an optimistic scenario where uncertainties are reduced by a factor of 5 and a correlation $\rho = 0.85$.}
    \label{fig:sensitivity-scalar}
\end{figure}
Following our statistical approach, using the experimental setups and NME scenarios described above, we estimate the sensitivity to $a_\nu$ at 90\% CL for a given scalar mass $m_S$. The result is shown in Fig.~\ref{fig:sensitivity-scalar}. The coloured blue, green and red regions at its top would be excluded by the existing searches at GERDA~II, KamLAND-Zen and NEMO-3, respectively. The maximally testable $m_S$ is determined by the isotope's $Q$ value and the minimum usable $T_\text{min}$ of the given experiment, $m_S < Q - T_\text{min}$. Thus, despite NEMO-3 having the smallest exposure, it still provides a relevant constraint due the large $Q$ value of $^{100}$Mo. As mentioned, for the future searches, we take a pessimistic and optimistic set of NMEs, corresponding to the blue, green and red bands for LEGEND-1000, nEXO and CUPID, respectively. In addition to the higher exposures, we also take a uniform value of $T_\text{min} = 100$~keV, e.g., extending the sensitivity of CUPID to $m_S \lesssim 2.9$~MeV. We note that the sensitivity to scalar emission $0\nu\beta\beta S$ (massive Majoron) was estimated in \cite{Brune:2018sab} but only taking into account the phase space suppression of the total $0\nu\beta\beta S$ decay rate due to a massive scalar. We here instead include the spectrum shape as well. Specifically, as $m_S$ increases, the phase space is suppressed and the $0\nu\beta\beta S$ spectrum shifts to lower electron energies, becoming more similar to SM $2\nu\beta\beta$ decay, see Fig.~\ref{fig:spectra}. We thus see that our sensitivity worsens more rapidly for increasing $m_S$ until $m_S$ becomes large enough for the spectrum to peak to the left of the SM $2\nu\beta\beta$ spectrum. Finally, we note that the PandaX collaboration has experimentally constrained $0\nu\beta\beta S$ emission in this scenario using $^{125}$Xe double beta decay data~\cite{PandaX:2025tls}.


\subsection{\texorpdfstring{Model I$^\prime$}{Model I'}: Coupling to neutrinos only}

We first extend the simple scalar emission case by allowing the scalar to decay to active neutrinos. This corresponds to Model I$^\prime$ (c.f., Sec.~\ref{sec:modelIprime}), where the coupling to $\chi$ is set to zero. The measured electron energy spectrum can then be expressed as
\begin{align}
    \frac{d\Gamma(\mathbf{x})}{dT} = 
    \frac{d\Gamma^{2\nu}_\text{SM}}{dT}
    + \frac{d\Gamma^{2\nu}_S(\mathbf{x})}{dT}
    + \frac{d\Gamma^{2\nu}_\text{int}(\mathbf{x})}{dT},
\end{align}
where the $2\nu_S\beta\beta$ spectrum is determined using Eq.~\eqref{eq:2chibb} with $m_{\chi_1} = m_{\chi_2} = 0$, and including a symmetry factor of $1/2$. The interference term is given in Eq.~\eqref{eq:2nbbint}. In full generality, the model parameters are described by the exotic particle masses $m_S$ and the set of couplings $a_\nu, b_\nu$, $\mathbf{x} = (m_S, a_\nu, b_\nu)$. For simplicity and minimality, we take $b_\nu = 0$, i.e., we only consider the $a_\nu$ coupling through which $S$ is produced.

\begin{figure}[t!]
    \centering
    \includegraphics[width=0.8\textwidth]{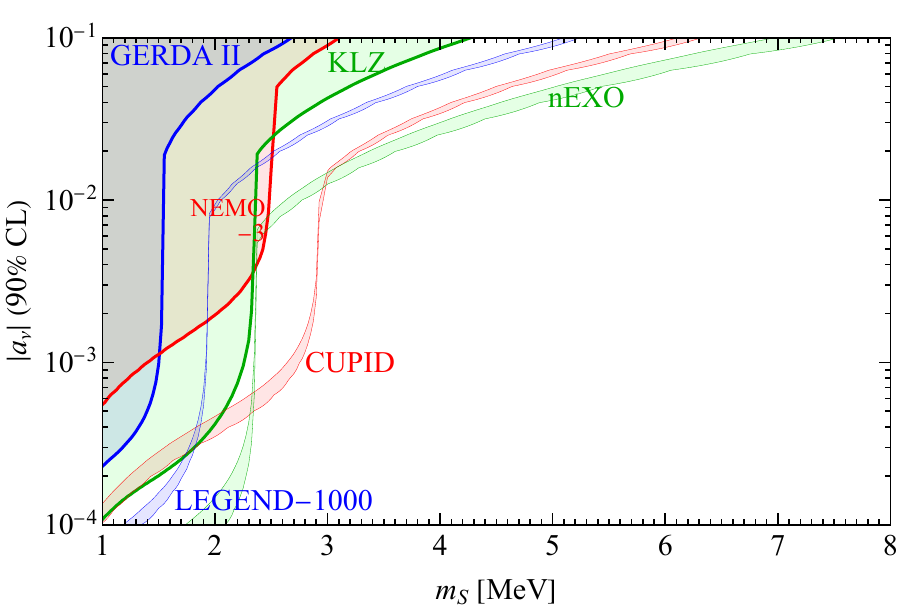}
    \caption{As Fig.~\ref{fig:sensitivity-scalar}, but showing the sensitivity on $a_\nu$ in Model I$^\prime$, with all couplings set to zero except for $a_\nu$.}
    \label{fig:sensitivity-nuonly}
\end{figure}
In Fig.~\ref{fig:sensitivity-nuonly}, we show the $a_\nu$ sensitivity as a function of $m_S$. The plot focuses on larger values of $m_S \geq 1$~MeV and $|a_\nu| \geq 10^{-4}$. For small $|a_\nu|$, with $m_S$ below the experimental threshold, the sensitivity is the same as for the scalar emission case in Fig.~\ref{fig:sensitivity-scalar}. This is because $0\nu\beta\beta S$ and $2\nu_S\beta\beta$ are largely indistinguishable as long as $S$ can be produced on-shell within the experimentally accessible electron spectrum and the total decay width $\Gamma_S$ remains small. On the other hand, above the threshold, $0\nu\beta\beta S$ produces no events whereas $2\nu_S\beta\beta$ is allowed. For larger $a_\nu$, the interference between SM $2\nu\beta\beta$ and $2\nu_S\beta\beta$ also becomes sizeable, scaling as $\propto a_\nu^2$. Thus, the sensitivity to $|a_\nu|$ extends both beyond the experimental threshold $m_S < Q - T_\text{min}$ and the ultimate $m_S$ threshold $m_S < Q$. For example, KamLAND-Zen provides a meaningful sensitivity, $|a_\nu| \lesssim 10^{-1}$ for masses up to $m_S \lesssim 4.3$~MeV. A future nEXO experiment may probe scalar masses up to $m_S \approx 7.5$~MeV.


\subsection{\texorpdfstring{Model I$^\prime$}{Model I'}: Dominant coupling to dark fermion}

We now extend the scope and explore the sensitivity where the scalar $S$ couples to a dark fermion $\chi$ as well. The measured electron energy spectrum now is
\begin{align}
    \frac{d\Gamma(\mathbf{x})}{dT} = 
    \frac{d\Gamma^{2\nu}_\text{SM}}{dT}
    + \frac{d\Gamma^{2\nu}_S(\mathbf{x})}{dT}
    + \frac{d\Gamma^{2\nu}_\text{int}(\mathbf{x})}{dT}
    + \frac{d\Gamma^{2\chi}_S(\mathbf{x})}{dT},
\end{align}
where the $2\chi_S\beta\beta$ spectrum is determined using Eq.~\eqref{eq:2chibb} with $m_{\chi_1} = m_{\chi_2} = m_\chi$, including a symmetry factor of $1/2$. In full generality, the model parameters are described by the exotic particle masses $m_S$, $m_\chi$ and the full set of couplings $a_\nu, b_\nu, a_\chi, b_\chi$, $\mathbf{x} = (m_S, m_\chi, a_\nu, b_\nu, a_\chi, b_\chi)$. For simplicity and to explore the full range by which the above scalar emission case can be modified, we take $b_\nu = b_\chi = 0$ and $a_\chi = 1$, i.e., we only keep the left-handed couplings and assume that $S$ couples dominantly to the dark fermion $\chi$. Because of this, $S$ almost exclusively decays to $\chi$, Br$(S\to\chi\chi) \approx 1$ for $a_\nu \lesssim 10^{-1}$ and sufficiently light $\chi$, $m_\chi \lesssim m_S/3$. Given these choices, we determine the 90\% CL constraint on $a_\nu$ for given masses $m_S$ and~$m_\chi$.

We note that for massless $\chi$, as considered below first, the model is equivalent to one where $S$ also couples to another active neutrino flavour, such as $\nu_\tau\nu_\tau$ and thus identifying $a_\chi \to a_\nu^{\tau\tau}$. Couplings to third generation neutrinos are constrained much more weakly and can be large \cite{Dev:2025tdv}.

\begin{figure}[t!]
    \centering
    \includegraphics[width=0.8\textwidth]{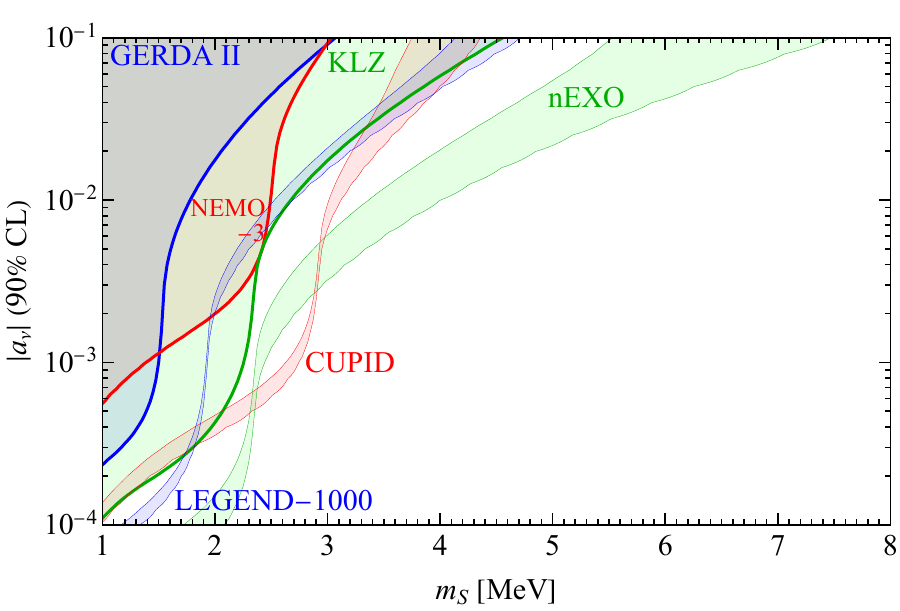}
    \caption{As Fig.~\ref{fig:sensitivity-scalar}, but showing the sensitivity on $a_\nu$ in Model I$^\prime$, with coupling $a_\chi = 1$, all other couplings set to zero and a massless $\chi$.}
    \label{fig:sensitivity-offshell}
\end{figure}
In Fig.~\ref{fig:sensitivity-offshell}, we show the $a_\nu$ sensitivity as a function of $m_S$, assuming a massless $\chi$. We again focus on the mass region near and above the $S$ production threshold. Qualitatively, the behaviour is equivalent to Fig.~\ref{fig:sensitivity-nuonly}, with sensitivities extending far above the $Q$ value. With $a_\chi = 1$, the $S$ decay width is much higher than in the previous case, resulting in broader finite width corrections and an improved sensitivity near the threshold. In future searches, this sensitivity will depend more strongly on the NME scenario considered, though. This is because the spectrum above the threshold is dominated by $2\chi_S\beta\beta$ production and not the interference with $2\nu_S\beta\beta$. The spectrum shape in this case, for $m_\chi = 0$, is largely indistinguishable from that of SM $2\nu\beta\beta$. The sensitivity is therefore strongly dependent on the ability to fix the overall spectrum normalization.

\begin{figure}[t!]
    \centering
    \includegraphics[width=0.8\textwidth]{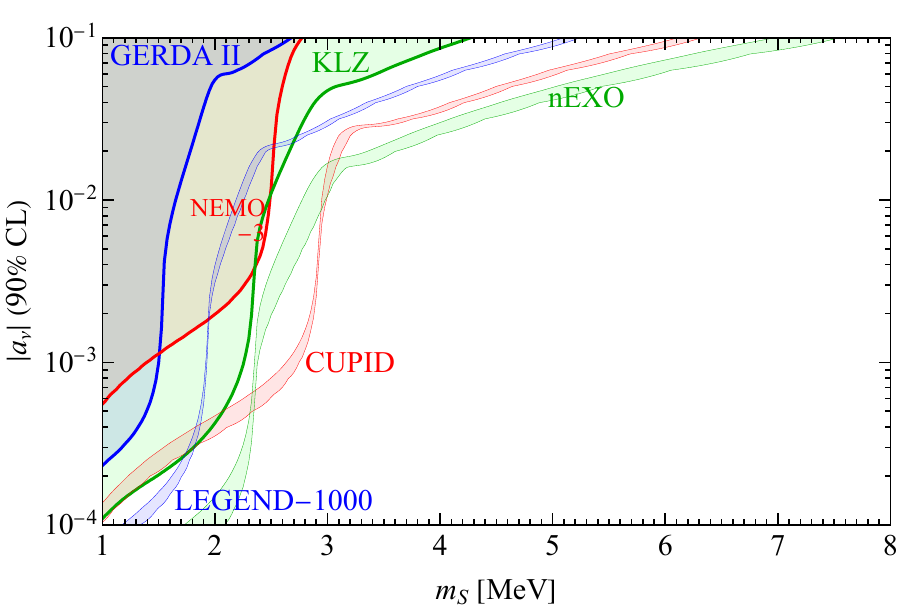}
    \caption{As Fig.~\ref{fig:sensitivity-scalar}, but showing the sensitivity on $a_\nu$ in Model I$^\prime$, with coupling $a_\chi = 1$, all other couplings set to zero and a massive $\chi$, $m_\chi = m_S / 3$.}
    \label{fig:sensitivity-offshell-massive}
\end{figure}
Of course, this increased sensitivity on $a_\nu$ is dependent on the choice of the other parameters of the model. This can be captured by an extended $\chi^2$ analysis incorporating two or more model parameters at a time. This is beyond the scope of this work but in Fig.~\ref{fig:sensitivity-offshell-massive}, we show the equivalent plot for $m_\chi = m_S / 3$ instead of a massless $\chi$. The overall effect is reduced due to the phase-space suppression from $m_\chi$, especially near the visible $m_S$ threshold, $T \gtrsim Q - T_\text{min}$. For large $m_S$, where $\chi$ cannot be produced on-shell either, the light neutrino emission $2\nu_S\beta\beta$ becomes dominant and the sensitivities approach that of~Fig.~\ref{fig:sensitivity-nuonly}.


\subsection{Effective operator and dependence on dark fermion mass}

So far, we have mainly discussed the sensitivity as a function of the scalar mediator mass, for a fixed or correlated dark fermion mass $m_\chi$. To illustrate the dependence on $m_\chi$, we consider the production of $\chi$ for a heavy scalar, $m_S \gg Q$, or, equivalently, via an effective operator. Specifically, and in order to explore the largest mass range, we look at the associated production of a massless fermion (active $\nu$ or a massless/very light $\chi$) together with a massive dark fermion, described by Eq.~\eqref{eq:2chibb-heavy},
\begin{align}
    \frac{d\Gamma(\mathbf{x})}{dT} = 
    \frac{d\Gamma^{2\nu}_\text{SM}}{dT}
    + \frac{d\Gamma^{2\chi}_\text{EFT}(\mathbf{x})}{dT}.
\end{align}
Here, we take $m_{\chi_1} = 0$, $m_{\chi_2} = m_\chi$. We determine the sensitivity on the effective operator strength $|\Lambda| = (|\Lambda_a^{-2}|^2 + \Lambda_b^{-2}|^2)^{-1/4}$ as a function of $m_\chi$, $\mathbf{x} = (m_\chi, |\Lambda|)$. 

In the context of our three models in Sec.~\ref{sec:UV}, such a scenario can be realized in Model~I with a dominant coupling $\overline{\nu^c} P_L \chi S^*$, either through a direct coupling $a_{\nu\chi}$ or large active-sterile mixing $\theta$. In this case, we have a mixed production of an active neutrino with a dark fermion, $\nu\chi_\text{EFT}\beta\beta$ when $S$ is heavy. It can also be realized in Model~II, taking, say, $\chi_1$ to be massless or very light, $m_{\chi_1} \ll Q$.

\begin{figure}[t!]
    \centering
    \includegraphics[width=0.8\textwidth]{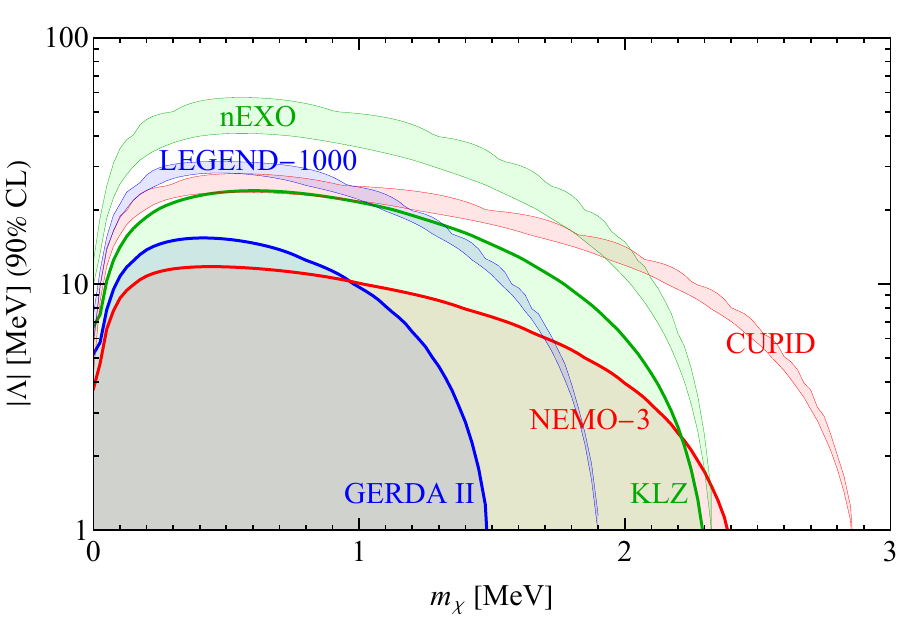}
    \caption{Sensitivity on the operator scale $\Lambda$ as a function of $m_\chi$, for $\nu\chi_{S}\beta\beta$ production of a single, massive dark fermion via a heavy $s$-channel mediator.}
    \label{fig:sensitivity-eft}
\end{figure}
The sensitivity on the EFT scale $\Lambda$ is shown in Fig.~\ref{fig:sensitivity-eft}, as a function of the dark fermion mass $m_\chi$. For $m_\chi \to 0$, we recover the scenario in Fig.~\ref{fig:sensitivity-offshell} for large $m_S$ and the associated sensitivity. Because the decay spectrum is largely identical to SM $2\nu\beta\beta$ decay, it is not the most easily testable. Instead, a finite $m_\chi$ leads to a pronounced kink in the spectrum. For large $m_\chi$, the decay will be phase-space suppressed and the optimal sensitivity is achieved for $m_\chi \approx 0.3 - 0.6$~MeV. The current limit on $|\Lambda|$ is largely set by KamLAND-Zen, up to $|\Lambda| \approx 20$~MeV. This may be improved to $|\Lambda| \approx 60$~MeV in the future. The use of $^{100}$Mo as isotope will have the benefit of probing heavier $\chi$ masses. While the EFT treatment requires, strictly speaking, $|\Lambda| \gg Q$ (or, rather, $m_S \gg Q$), it is applicable for fairly light mediator masses, not too far above the $Q$-value threshold.


\section{Conclusion}
\label{sec:conclusion}

Double beta decay processes are sensitive avenues to probe physics beyond the SM. While the neutrinoless mode, $0\nu\beta\beta$ is the key process to probe the Majorana nature of neutrinos, one can also expect distinct signatures over the SM $2\nu\beta\beta$ decay spectrum, in presence of new light particles. In this work, we have explored a class of scenarios where a light scalar, $S$, couples both to active neutrinos and, potentially, a new fermionic particle $\chi$. Owing to these new interactions, on- and off-shell production of $S$ followed by its transition into neutrinos or exotic fermions, a spectral distortion in the double beta decay spectrum can be observed. We have constructed simplified model scenarios responsible for such exotic double beta decay contributions. Focusing on scalar $s$-channel mediation, we considered two Dirac neutrino scenarios (Model I and I$^\prime$) differing by the presence or absence of $\nu$–$\chi$ mixing, the latter enforced through a stabilizing $\mathcal{Z}_2$ symmetry that makes $\chi$ a potential dark matter candidate. We further introduced a third model (Model II) involving both Dirac and Majorana exotic fermions to illustrate the broader range of possible interactions. 

We analyzed how the double beta decay spectrum is modified by the presence of the mediator scalar $S$ and the exotic fermion species $\chi$, highlighting the dependence of the spectral shape on their masses and couplings. For massless $\chi$, we showed that when $S$ is sufficiently light to be produced on-shell, the resulting $2\chi_S\beta\beta$ spectrum closely follows the scalar emission shape, with only minor deviations arising from finite-width effects of the mediating scalar. As $m_S$ approaches the kinematic threshold or becomes too heavy to be produced on-shell, the spectrum shifts toward lower electron energies or enters an off-shell regime in which the dominant contributions move to larger $T$, eventually matching the SM $2\nu\beta\beta$ distribution in the $m_S \gg Q$ limit. Introducing non-zero $\chi$ masses further restricts on-shell production of the mediator and suppresses visible distortions in the electron spectrum, making heavier $\chi$ scenarios increasingly challenging to probe. Finally, in the case where $S$ couples exclusively to neutrinos, we demonstrated that the full double beta decay spectrum receives both on-shell and off-shell scalar-mediated contributions, including destructive interference with the SM amplitude, leading to a characteristic spectral suppression near the threshold. These results collectively illustrate the diverse range of spectral signatures that light scalars and exotic fermions can imprint on double beta decay observables.

\begin{figure}[t!]
    \centering
    \includegraphics[width=0.7\textwidth]{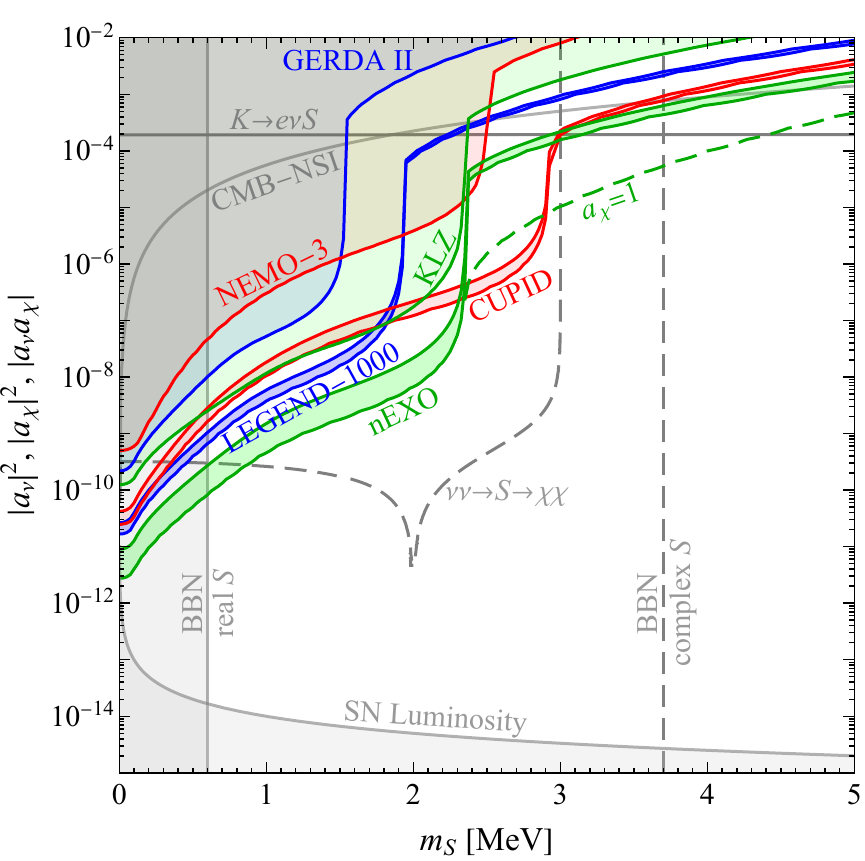}
    \caption{Summary of double beta decay sensitivities as discussed in Sec.~\ref{sec:results} compared with the most relevant CMB-NSI, BBN, Supernovae and kaon constraints, c.f., Sec.~\ref{sec:DM}.}
\label{fig:summary_plot}
\end{figure}
In Model I$^\prime$, the $\chi$ can be considered as a potential dark matter candidate (or a component of the dark sector) due to the absence of mixing with the active neutrino sector, due to a $\mathcal{Z}_2$ symmetry. Here, we discussed cosmological and laboratory constraints. Collisional damping, CMB, and BBN limits collectively require lower bounds on light DM and mediator masses, while SN data, rare kaon decays further restrict the viable couplings for larger mediator scalar mass. Together, these complementary constraints significantly narrow the parameter space in which the light scalar–fermion dark sector remains consistent with current observations, as shown in Fig.~\ref{fig:cosmo_lab}. 

Incorporating statistical, experimental and theoretical uncertainties on the double beta decay spectrum, we have estimated the sensitivity of a set of current and future double beta decay experiments towards exotic scalar emission $0\nu\beta\beta S$ as well as massless and massive exotic fermion production from the mediator scalar decay $2\chi_S\beta\beta$. This is summarized in Fig.~\ref{fig:summary_plot}, comparing the double beta decay constraints and future sensitivity with the most relevant cosmological and laboratory bounds. The darker blue, green and red solid curves illustrate the exclusion bounds on $|a_\nu|$ in Model~I$^\prime$ for $a_\chi = 0$, derived from the three current experimental searches, GERDA II, KamLAND-Zen and NEMO-3, respectively, at 90\% CL. Likewise, the bands indicate the projected sensitivities of the future experiments LEGEND-1000, nEXO and CUPID, using different nuclear matrix element uncertainties. In contrast, the scenario with $a_\chi = 1$ coupled to a massless dark fermion (or another generation of active neutrino such as $\nu_\tau$) is represented by the dashed curve for nEXO as the most sensitive future search. While BBN, SN data and kaon decay bounds impose stringent bounds, future double beta decay experiment can probe the parameter space for $m_S \gtrsim 0.6$~MeV and couplings $10^{-5} \lesssim |a_\nu| \lesssim 10^{-2}$ for a real scalar $S$. While not visible in Fig.~\ref{fig:summary_plot}, scalars lighter than $m_S \lesssim 0.1$~keV and coupling with $a_\nu \lesssim 10^{-5}$ will not be thermally produced by the time of BBN and the corresponding bound will not apply, see Fig.~\ref{fig:cosmo_lab}. Current double beta decay experiments, as shown in Fig.~\ref{fig:summary_plot} and recently reported by PandaX \cite{PandaX:2025tls} are already probing this regime and future searches will extend it to $|a_\nu| \approx 2\times 10^{-6}$. Double beta decay experiments thus provide the most stringent laboratory bounds and have the potential to probe light scalars coupling to neutrinos as well as to light fermions in a dark sector, extending beyond the double beta decay $Q$ values. The experimental sensitivities are in a parameter space relevant for cosmological considerations, providing independent, laboratory tests of light exotic physics.


\subsection*{Note}
As we were finalizing this manuscript, we became aware of an independent experimental analysis by the PandaX collaboration on a similar topic \cite{PandaX:2025tls}.

\acknowledgments
F.~F.~D. would like to thank Jianglai Liu, Shao-Feng Ge and Oleg Titov for useful discussion. A.~H.-B. would like to thank Martin Hirsch for useful discussions. F.~F.~D. and N.-I.~B. acknowledge support from the UK Science and Technology Facilities Council (STFC) via the Consolidated Grant ST/X000613. F.~F.~D. would like to thank the Tsung-Dao Lee Institute, where part of this work was undertaken, for their hospitality. C.~M. acknowledges the support from the Royal Society, UK, through the Newton International Fellowship (grant number NIF$\backslash$R1$\backslash$221737). The work of A.~H.-B. is supported by the Spanish grants PID2023-147306NB-I00 and CEX2023-001292-S (MCIU/AEI/10.13039/501100011033), as well as CIPROM/2021/054, CIACIF/2021/100 and CIBEFP/2023/090 (Generalitat Valenciana). S.~S. would like to thank University College London for their hospitality during part of this work.


\appendix

\section{Normalization of \texorpdfstring{$\beta\beta$}{beta beta} decay widths}
\label{App:normalization}

The nuclear structure appearing in our processes, i.e., $\chi_1 \chi_2 \beta \beta$ and Majoron decay, is the same as that involved in ordinary neutrinoless double beta decay. Therefore, we can make use of the NMEs of that process. To do so, our normalization must be matched to that used in the literature in order to consistently employ the numerical results for $\mathcal{M}_{0\nu}$.

Within our approach, it is straightforward to recover the standard results for $0\nu\beta\beta$ decay. The process amplitude can be factorized into leptonic and hadronic contributions,
\begin{equation}
    \mathcal{A}_{0\nu} = \sum_n L_{\mu\nu} H^{\mu\nu},
\end{equation}
where $L_{\mu\nu}$ and $H^{\mu\nu}$ denote the leptonic and hadronic contributions, respectively, and the sum over $n$ accounts for the nuclear intermediate states. Assuming the closure approximation, the two decouple and, using the symmetry under $\mu \leftrightarrow \nu$ in the hadronic sector, the leptonic part can be written as
\begin{equation}
    L_{\mu \nu} = 
    \mathcal{N}_{0\nu} m_{\beta\beta} g_{\mu \nu} \bar{u}_1 P_R v_2,
\end{equation}
where $\mathcal{N}_{0\nu}$ is a normalization factor chosen to match the standard results in the literature, and $m_{\beta\beta}$ is the effective double beta decay neutrino mass. The corresponding decay width is then expressed as
\begin{equation}
    \Gamma_{0\nu} = 
    \left|\frac{m_{\beta\beta}}{m_e} \right|^2 
    \left| \mathcal{M}_{0\nu} \right|^2 \mathcal{G}_{0\nu}.
\end{equation}
where $\mathcal{G}_{0\nu}$ and $\mathcal{M}_{0\nu}$ denote the phase space factor and the NME, respectively. The phase space factor can be computed as
\begin{align}
    \mathcal{G}_{0\nu} 
    &= \left|\mathcal{N}_{0\nu}\right|^2 
    \int \prod_{i=1}^2 \frac{d^3 \vec{p}_i}{(2\pi)^3 2E_i} 2\, p_1 \cdot p_2  F_0(p_1) F_0(p_2) \delta(Q - T) \nonumber\\
    &= \frac{2\left| \mathcal{N}_{0\nu} \right|^2}{(2\pi)^4} \int_{m_e}^{Q + m_e } |\vec{p}_1| |\vec{p}_2| E_1 E_2 F_0(p_1) F_0(p_2)  dE_1,
\end{align}
where $F_0(p)$ is the Fermi function, and we have used $E_2 = Q + 2 m_e - E_1$ and $|\vec{p}_i| = \sqrt{E_i^2 - m_i^2}$. Comparing with the standard result in the literature, we obtain
\begin{equation}
    \left|\mathcal{N}_{0\nu} \right|^2 
    = \frac{G_F^4 \cos^4\theta_c}{4\pi R^2},
\end{equation}
in such a way the decay width is given by
\begin{equation}
    \Gamma_{0 \nu} = 
    \left| \frac{m_{\beta\beta}}{m_e} \right|^2 
    \frac{G_F^4 \cos^4(\theta_c) m_e^2}{32 \pi^5 R^2} 
    \left| \mathcal{M}_{0 \nu} \right|^2 
    \int_{m_e}^{Q + m_e } |\vec{p}_1| |\vec{p}_2| E_1 E_2 F_0(p_1) F_0(p_2)  dE_1,
\end{equation}
and reproduce the known result.

For the case where $\chi_1$ and $\chi_2$ are antineutrinos, one must also account for the interference with standard $2\nu \beta \beta$. Following a similar approach, the leptonic contribution can be written as
\begin{equation}
    L_{\mu\nu} 
    = \mathcal{N}_{2\nu} g_{\mu\nu} 
    \left[\bar{u}_1 P_R v_2\right] 
    \left[\bar{u}_4 P_R v_3\right],
\end{equation}
where $p_1$ and $p_2$ correspond to the electron momenta, while $p_3$ and $p_4$ correspond to the neutrinos. The decay width is
\begin{equation}
    \Gamma_{2\nu} = \left|\mathcal{M}_{2\nu}\right|^2 \mathcal{G}_{2\nu},
\end{equation}
with the phase space factor
\begin{align}
    \mathcal{G}_{2 \nu} 
    &= \left|\mathcal{N}_{2\nu}\right|^2 
    \int \prod_{i=1}^4 \frac{d^3 \vec{p}_i}{(2\pi)^3 2E_i} 4\, p_1 \cdot p_2 
    F_0(p_1) F_0(p_2) p_3 \cdot p_4 \delta(Q - T - E_3 - E_4) \nonumber\\
    &= \frac{4\left|\mathcal{N}_{2\nu}\right|^2}{(2\pi)^8} 
    \int \frac{(Q - T)^5}{30} |\vec{p}_1| |\vec{p}_2| E_1 E_2 F_0(p_1) F_0(p_2) 
    dE_1 dE_2.
\end{align}
Comparison with the standard literature then gives
\begin{equation}
    \left|\mathcal{N}_{2\nu}\right|^2 
    = 8\pi\frac{G_F^4 \cos^4(\theta_c)}{m_e^2}.
\end{equation}
Thus, the decay width is given by
\begin{equation}
    \Gamma_{2\nu} 
    = \frac{G_F^4 \cos^4(\theta_c)}{8 \pi^7m_e^2}
    \left|\mathcal{M}_{2\nu}\right|^2 
    \int \frac{(Q - T)^5}{30} |\vec{p}_1| |\vec{p}_2| E_1 E_2 F_0(p_1) F_0(p_2) 
    dE_1 dE_2,
\end{equation}
recovering the known result.


\section{Calculation of scattering rates}
\label{sec:appendix}

The squared amplitude, summed over final and averaged over initial spins, of DM $\chi$ to neutrino annihilation via $s$-channel scalar mediation, $\chi (p_1) + \chi(p_2) \to S \to \nu (p_3) + \nu(p_4)$, can be written as
\begin{align}
    |\mathcal{M}_s|^2 = 
    \frac{1}{(s-m_S^2)^2 + m_S^2\Gamma_S^2} 
    \left[(|a_\chi|^2 + |b_\chi|^2)(p_1\cdot p_2) - 2m_\chi^2 
    \text{Re}(a_\chi b^*_\chi)\right]
    \left[(|a_\nu|^2+|b_\nu|^2)(p_3\cdot p_4)\right],
\end{align}
where $p_1\cdot p_2 = (s-2m_\chi^2)/2$, $p_3\cdot p_4 = s/2$. Following \cite{Wells:1994qy, Olivares-DelCampo:2017feq}, we have estimated the thermally averaged annihilation cross section $\langle \sigma v_{\rm{rel}}\rangle$ which will be constrained by present-day relic density criteria for DM candidates in the universe, where we have considered $v_{\rm{rel}} = 10^{-3} \ll c$. 

Similarly, the squared amplitude for elastic scattering between DM $\chi$ and neutrinos, $\chi(p_1) + \bar{\nu} (p_2) \to \bar{\chi} (p_3) + \nu(p_4)$ via $t$-channel scalar mediation can be expressed as
\begin{align}
    |\mathcal{M}_t|^2 = 
    \frac{1}{2(t-m_S^2)^2} 
    [(p_3\cdot p_1)(|a_\chi|^2+|b_\chi|^2) + (p_4\cdot p_2)(|a_\nu|^2+|b_\nu|^2) 
    - 2m_\chi^2 \text{Re}(a_\chi b_\chi^*)],
\end{align}
where $p_3\cdot p_1 = (t-2m_\chi^2)/2$ and $p_4\cdot p_2 = t/2$. Using this, the elastic scattering cross section can be estimated as $\sigma_\text{el} = 1/(16\pi s)\sum_\text{spins} |\mathcal{M}_t|^2$ and this is constrained by the bounds from collisional damping \cite{Boehm:2000gq, Boehm:2001hm, Boehm:2004th}.

Furthermore, we have obtained the scattering cross section for $\nu (p_1) + \nu (p_2) \to \chi (p_3) + \chi (p_4)$ via $s$-channel scalar mediation,
\begin{align}
    \sigma(\nu\nu\to S\to\chi\chi) &= 
    \frac{1}{32\pi s \left[(s-m_S^2)^2 + m_S^2 \Gamma_S^2 \right]}  
    \sqrt{1 - \frac{4m_\chi^2}{s}} \nonumber\\
    &\times
    \left[(|a_\nu|^2+|b_\nu|^2)(p_1\cdot p_2)\right]
    \left[(|a_\chi|^2+|b_\chi|^2)(p_3\cdot p_4)-2m_\chi^2 
    \text{Re}(a_\chi b_\chi^*)\right],
    \label{eq:nunuchichi_CS}
\end{align}
with $p_1\cdot p_2 = s/2$ and $p_3\cdot p_4 = (s-2m_\chi^2)/2$.


\bibliographystyle{utcaps_mod}
\bibliography{2nbb}

@article{Barabash:2019nnr,
    author = "Barabash, A. S.",
    editor = "Civitarese, Osvaldo and Stekl, Ivan and Suhonen, Jouni",
    title = "{Average and recommended half-life values for two-neutrino double beta decay: upgrade-2019}",
    eprint = "1907.06887",
    archivePrefix = "arXiv",
    primaryClass = "nucl-ex",
    doi = "10.1063/1.5130963",
    journal = "AIP Conf. Proc.",
    volume = "2165",
    number = "1",
    pages = "020002",
    year = "2019"
}

@article{Deppisch:2012nb,
    author = "Deppisch, Frank F. and Hirsch, Martin and Pas, Heinrich",
    title = "{Neutrinoless Double Beta Decay and Physics Beyond the Standard Model}",
    eprint = "1208.0727",
    archivePrefix = "arXiv",
    primaryClass = "hep-ph",
    reportNumber = "IFIC-12-56",
    doi = "10.1088/0954-3899/39/12/124007",
    journal = "J. Phys. G",
    volume = "39",
    pages = "124007",
    year = "2012"
}

@article{Graf:2018ozy,
    author = "Graf, Lukas and Deppisch, Frank F. and Iachello, Francesco and Kotila, Jenni",
    title = "{Short-Range Neutrinoless Double Beta Decay Mechanisms}",
    eprint = "1806.06058",
    archivePrefix = "arXiv",
    primaryClass = "hep-ph",
    doi = "10.1103/PhysRevD.98.095023",
    journal = "Phys. Rev. D",
    volume = "98",
    number = "9",
    pages = "095023",
    year = "2018"
}

@article{Cirigliano:2018yza,
    author = "Cirigliano, V. and Dekens, W. and de Vries, J. and Graesser, M. L. and Mereghetti, E.",
    title = "{A neutrinoless double beta decay master formula from effective field theory}",
    eprint = "1806.02780",
    archivePrefix = "arXiv",
    primaryClass = "hep-ph",
    reportNumber = "LA-UR-18-24895, Nikhef 2018-023, NIKHEF-2018-023, DESY-18-072",
    doi = "10.1007/JHEP12(2018)097",
    journal = "JHEP",
    volume = "12",
    pages = "097",
    year = "2018"
}

@article{NEMO-3:2009fxe,
    author = "Argyriades, J. and others",
    collaboration = "NEMO-3",
    title = "{Measurement of the two neutrino double beta decay half-life of Zr-96 with the NEMO-3 detector}",
    eprint = "0906.2694",
    archivePrefix = "arXiv",
    primaryClass = "nucl-ex",
    doi = "10.1016/j.nuclphysa.2010.07.009",
    journal = "Nucl. Phys. A",
    volume = "847",
    pages = "168--179",
    year = "2010"
}

@article{NEMO-3:2016qxo,
    author = "Arnold, R. and others",
    collaboration = "NEMO-3",
    title = "{Measurement of the 2$\nu\beta\beta$ decay half-life of $^{150}$Nd and a search for 0$\nu\beta\beta$ decay processes with the full exposure from the NEMO-3 detector}",
    eprint = "1606.08494",
    archivePrefix = "arXiv",
    primaryClass = "hep-ex",
    doi = "10.1103/PhysRevD.94.072003",
    journal = "Phys. Rev. D",
    volume = "94",
    number = "7",
    pages = "072003",
    year = "2016"
}

@article{NEMO-3:2016mvr,
    author = "Arnold, R. and others",
    collaboration = "NEMO-3",
    title = "{Measurement of the double-beta decay half-life and search for the neutrinoless double-beta decay of $^{48}$Ca with the NEMO-3 detector}",
    eprint = "1604.01710",
    archivePrefix = "arXiv",
    primaryClass = "hep-ex",
    doi = "10.1103/PhysRevD.93.112008",
    journal = "Phys. Rev. D",
    volume = "93",
    number = "11",
    pages = "112008",
    year = "2016"
}

@article{Arnold:2018tmo,
    author = "Arnold, R. and others",
    title = "{Final results on $^{82}{Se}$ double beta decay to the ground state of $^{82}{Kr}$ from the NEMO-3 experiment}",
    eprint = "1806.05553",
    archivePrefix = "arXiv",
    primaryClass = "hep-ex",
    doi = "10.1140/epjc/s10052-018-6295-x",
    journal = "Eur. Phys. J. C",
    volume = "78",
    number = "10",
    pages = "821",
    year = "2018"
}

@article{Simkovic:2018rdz,
    author = "\v{S}imkovic, Fedor and Dvornick\'y, Rastislav and Stef\'anik, Du\v{s}a\v{n} and Faessler, Amand",
    title = "{Improved description of the $2\nu\beta\beta$ -decay and a possibility to determine the effective axial-vector coupling constant}",
    eprint = "1804.04227",
    archivePrefix = "arXiv",
    primaryClass = "nucl-th",
    doi = "10.1103/PhysRevC.97.034315",
    journal = "Phys. Rev. C",
    volume = "97",
    number = "3",
    pages = "034315",
    year = "2018"
}

@article{Chikashige:1980ui,
    author = "Chikashige, Y. and Mohapatra, Rabindra N. and Peccei, R. D.",
    title = "{Are There Real Goldstone Bosons Associated with Broken Lepton Number?}",
    reportNumber = "MPI-PAE-PTH-36-80",
    doi = "10.1016/0370-2693(81)90011-3",
    journal = "Phys. Lett. B",
    volume = "98",
    pages = "265--268",
    year = "1981"
}

@article{Gelmini:1980re,
    author = "Gelmini, G. B. and Roncadelli, M.",
    title = "{Left-Handed Neutrino Mass Scale and Spontaneously Broken Lepton Number}",
    reportNumber = "MPI-PAE-PTH-50-80",
    doi = "10.1016/0370-2693(81)90559-1",
    journal = "Phys. Lett. B",
    volume = "99",
    pages = "411--415",
    year = "1981"
}

@article{Carone:1993jv,
    author = "Carone, Christopher D.",
    title = "{Double beta decay with vector majorons}",
    eprint = "hep-ph/9302290",
    archivePrefix = "arXiv",
    reportNumber = "HUTP-93-A007",
    doi = "10.1016/0370-2693(93)90605-H",
    journal = "Phys. Lett. B",
    volume = "308",
    pages = "85--88",
    year = "1993"
}

@article{Bamert:1994hb,
    author = "Bamert, P. and Burgess, C. P. and Mohapatra, R. N.",
    title = "{Multi - Majoron modes for neutrinoless double beta decay}",
    eprint = "hep-ph/9412365",
    archivePrefix = "arXiv",
    reportNumber = "NEIP-94-010, MCGILL-94-18A, UMD-PP-95-78",
    doi = "10.1016/0550-3213(95)00273-U",
    journal = "Nucl. Phys. B",
    volume = "449",
    pages = "25--48",
    year = "1995"
}

@article{Hirsch:1995in,
    author = "Hirsch, M. and Klapdor-Kleingrothaus, H. V. and Kovalenko, S. G. and Pas, H.",
    title = "{On the observability of majoron emitting double beta decays}",
    eprint = "hep-ph/9511227",
    archivePrefix = "arXiv",
    doi = "10.1016/0370-2693(96)00038-X",
    journal = "Phys. Lett. B",
    volume = "372",
    pages = "8--14",
    year = "1996"
}

@article{Blum:2018ljv,
    author = "Blum, Kfir and Nir, Yosef and Shavit, Michal",
    title = "{Neutrinoless double-beta decay with massive scalar emission}",
    eprint = "1802.08019",
    archivePrefix = "arXiv",
    primaryClass = "hep-ph",
    doi = "10.1016/j.physletb.2018.08.022",
    journal = "Phys. Lett. B",
    volume = "785",
    pages = "354--361",
    year = "2018"
}

@article{Cepedello:2018zvr,
    author = "Cepedello, Ricardo and Deppisch, Frank F. and Gonz\'alez, Lorena and Hati, Chandan and Hirsch, Martin",
    title = "{Neutrinoless Double-$\beta$ Decay with Nonstandard Majoron Emission}",
    eprint = "1811.00031",
    archivePrefix = "arXiv",
    primaryClass = "hep-ph",
    reportNumber = "IFIC/18-38",
    doi = "10.1103/PhysRevLett.122.181801",
    journal = "Phys. Rev. Lett.",
    volume = "122",
    number = "18",
    pages = "181801",
    year = "2019"
}

@article{Deppisch:2020mxv,
    author = "Deppisch, Frank F. and Graf, Lukas and \v{S}imkovic, Fedor",
    title = "{Searching for New Physics in Two-Neutrino Double Beta Decay}",
    eprint = "2003.11836",
    archivePrefix = "arXiv",
    primaryClass = "hep-ph",
    doi = "10.1103/PhysRevLett.125.171801",
    journal = "Phys. Rev. Lett.",
    volume = "125",
    number = "17",
    pages = "171801",
    year = "2020"
}

@article{Berezinsky:1993fm,
    author = "Berezinsky, V. and Valle, J. W. F.",
    title = "{The KeV majoron as a dark matter particle}",
    eprint = "hep-ph/9309214",
    archivePrefix = "arXiv",
    reportNumber = "FTUV-93-35, LNGS-93-79",
    doi = "10.1016/0370-2693(93)90140-D",
    journal = "Phys. Lett. B",
    volume = "318",
    pages = "360--366",
    year = "1993"
}

@article{Garcia-Cely:2017oco,
    author = "Garcia-Cely, Camilo and Heeck, Julian",
    title = "{Neutrino Lines from Majoron Dark Matter}",
    eprint = "1701.07209",
    archivePrefix = "arXiv",
    primaryClass = "hep-ph",
    reportNumber = "ULB-TH-17-01",
    doi = "10.1007/JHEP05(2017)102",
    journal = "JHEP",
    volume = "05",
    pages = "102",
    year = "2017"
}

@article{Brune:2018sab,
    author = {Brune, Tim and P\"as, Heinrich},
    title = "{Massive Majorons and constraints on the Majoron-neutrino coupling}",
    eprint = "1808.08158",
    archivePrefix = "arXiv",
    primaryClass = "hep-ph",
    reportNumber = "DO-TH 18/23",
    doi = "10.1103/PhysRevD.99.096005",
    journal = "Phys. Rev. D",
    volume = "99",
    number = "9",
    pages = "096005",
    year = "2019"
}

@article{Deppisch:2020sqh,
    author = "Deppisch, Frank F. and Graf, Lukas and Rodejohann, Werner and Xu, Xun-Jie",
    title = "{Neutrino Self-Interactions and Double Beta Decay}",
    eprint = "2004.11919",
    archivePrefix = "arXiv",
    primaryClass = "hep-ph",
    doi = "10.1103/PhysRevD.102.051701",
    journal = "Phys. Rev. D",
    volume = "102",
    number = "5",
    pages = "051701",
    year = "2020"
}

@article{ParticleDataGroup:2024cfk,
    author = "Navas, S. and others",
    collaboration = "Particle Data Group",
    title = "{Review of particle physics}",
    doi = "10.1103/PhysRevD.110.030001",
    journal = "Phys. Rev. D",
    volume = "110",
    number = "3",
    pages = "030001",
    year = "2024"
}

@article{Wells:1994qy,
    author = "Wells, James D.",
    title = "{Annihilation cross-sections for relic densities in the low velocity limit}",
    eprint = "hep-ph/9404219",
    archivePrefix = "arXiv",
    reportNumber = "UM-TH-94-10",
    month = "3",
    year = "1994"
}

@article{Olivares-DelCampo:2017feq,
    author = "Olivares-Del Campo, Andr\'es and B\oe{}hm, C\'eline and Palomares-Ruiz, Sergio and Pascoli, Silvia",
    title = "{Dark matter-neutrino interactions through the lens of their cosmological implications}",
    eprint = "1711.05283",
    archivePrefix = "arXiv",
    primaryClass = "hep-ph",
    reportNumber = "IFIC-17-54, IPPP-17-84",
    doi = "10.1103/PhysRevD.97.075039",
    journal = "Phys. Rev. D",
    volume = "97",
    number = "7",
    pages = "075039",
    year = "2018"
}

@article{Boehm:2000gq,
    author = "Boehm, C. and Fayet, Pierre and Schaeffer, R.",
    title = "{Constraining dark matter candidates from structure formation}",
    eprint = "astro-ph/0012504",
    archivePrefix = "arXiv",
    doi = "10.1016/S0370-2693(01)01060-7",
    journal = "Phys. Lett. B",
    volume = "518",
    pages = "8--14",
    year = "2001"
}

@article{Boehm:2001hm,
    author = "Boehm, Celine and Riazuelo, Alain and Hansen, Steen H. and Schaeffer, Richard",
    title = "{Interacting dark matter disguised as warm dark matter}",
    eprint = "astro-ph/0112522",
    archivePrefix = "arXiv",
    reportNumber = "SACLAY-SPH-T-01-147",
    doi = "10.1103/PhysRevD.66.083505",
    journal = "Phys. Rev. D",
    volume = "66",
    pages = "083505",
    year = "2002"
}

@article{Agostini:2022zub,
    author = "Agostini, Matteo and Benato, Giovanni and Detwiler, Jason A. and Men{\'e}ndez, Javier and Vissani, Francesco",
    title = "{Toward the discovery of matter creation with neutrinoless {\ensuremath{\beta}}{\ensuremath{\beta}} decay}",
    eprint = "2202.01787",
    archivePrefix = "arXiv",
    primaryClass = "hep-ex",
    doi = "10.1103/RevModPhys.95.025002",
    journal = "Rev. Mod. Phys.",
    volume = "95",
    number = "2",
    pages = "025002",
    year = "2023"
}

@article{Boehm:2004th,
    author = "Boehm, Celine and Schaeffer, Richard",
    title = "{Constraints on dark matter interactions from structure formation: Damping lengths}",
    eprint = "astro-ph/0410591",
    archivePrefix = "arXiv",
    doi = "10.1051/0004-6361:20042238",
    journal = "Astron. Astrophys.",
    volume = "438",
    pages = "419--442",
    year = "2005"
}

@article{Akita:2023yga,
    author = "Akita, Kensuke and Ando, Shin'ichiro",
    title = "{Constraints on dark matter-neutrino scattering from the Milky-Way satellites and subhalo modeling for dark acoustic oscillations}",
    eprint = "2305.01913",
    archivePrefix = "arXiv",
    primaryClass = "astro-ph.CO",
    reportNumber = "CTPU-PTC-23-16",
    doi = "10.1088/1475-7516/2023/11/037",
    journal = "JCAP",
    volume = "11",
    pages = "037",
    year = "2023"
}

@article{Dev:2025tdv,
    author = "Dev, P. S. Bhupal and Kim, Doojin and Sathyan, Deepak and Sinha, Kuver and Zhang, Yongchao",
    title = "{New Constraints on Neutrino-Dark Matter Interactions: A Comprehensive Analysis}",
    eprint = "2507.01000",
    archivePrefix = "arXiv",
    primaryClass = "hep-ph",
    reportNumber = "CETUP-2023-022",
    month = "7",
    year = "2025"
}

@article{Boehm:2013jpa,
    author = "Boehm, C{\'e}line and Dolan, Matthew J. and McCabe, Christopher",
    title = "{A Lower Bound on the Mass of Cold Thermal Dark Matter from Planck}",
    eprint = "1303.6270",
    archivePrefix = "arXiv",
    primaryClass = "hep-ph",
    reportNumber = "IPPP-13-18, DCPT-13-36",
    doi = "10.1088/1475-7516/2013/08/041",
    journal = "JCAP",
    volume = "08",
    pages = "041",
    year = "2013"
}

@article{Wilkinson:2016gsy,
    author = "Wilkinson, Ryan J. and Vincent, Aaron C. and B{\oe}hm, C{\'e}line and McCabe, Christopher",
    title = "{Ruling out the light weakly interacting massive particle explanation of the Galactic 511 keV line}",
    eprint = "1602.01114",
    archivePrefix = "arXiv",
    primaryClass = "astro-ph.CO",
    reportNumber = "IPPP-16-09",
    doi = "10.1103/PhysRevD.94.103525",
    journal = "Phys. Rev. D",
    volume = "94",
    number = "10",
    pages = "103525",
    year = "2016"
}

@article{Mangano:2001iu,
    author = "Mangano, G. and Miele, G. and Pastor, S. and Peloso, M.",
    title = "{A Precision calculation of the effective number of cosmological neutrinos}",
    eprint = "astro-ph/0111408",
    archivePrefix = "arXiv",
    reportNumber = "DSF-37-2001, MPI-PHT-2001-51",
    doi = "10.1016/S0370-2693(02)01622-2",
    journal = "Phys. Lett. B",
    volume = "534",
    pages = "8--16",
    year = "2002"
}

@article{Mangano:2005cc,
    author = "Mangano, Gianpiero and Miele, Gennaro and Pastor, Sergio and Pinto, Teguayco and Pisanti, Ofelia and Serpico, Pasquale D.",
    title = "{Relic neutrino decoupling including flavor oscillations}",
    eprint = "hep-ph/0506164",
    archivePrefix = "arXiv",
    reportNumber = "DSF-16-2005, IFIC-05-17, MPP-2005-36",
    doi = "10.1016/j.nuclphysb.2005.09.041",
    journal = "Nucl. Phys. B",
    volume = "729",
    pages = "221--234",
    year = "2005"
}

@article{Kolb:1986nf,
    author = "Kolb, Edward W. and Turner, Michael S. and Walker, Terrence P.",
    title = "{The Effect of Interacting Particles on Primordial Nucleosynthesis}",
    reportNumber = "FERMILAB-PUB-86-077-A",
    doi = "10.1103/PhysRevD.34.2197",
    journal = "Phys. Rev. D",
    volume = "34",
    pages = "2197",
    year = "1986"
}

@article{Serpico:2004nm,
    author = "Serpico, Pasquale Dario and Raffelt, Georg G.",
    title = "{MeV-mass dark matter and primordial nucleosynthesis}",
    eprint = "astro-ph/0403417",
    archivePrefix = "arXiv",
    reportNumber = "MPP-2004-34",
    doi = "10.1103/PhysRevD.70.043526",
    journal = "Phys. Rev. D",
    volume = "70",
    pages = "043526",
    year = "2004"
}

@article{Boehm:2012gr,
    author = "Boehm, Celine and Dolan, Matthew J. and McCabe, Christopher",
    title = "{Increasing Neff with particles in thermal equilibrium with neutrinos}",
    eprint = "1207.0497",
    archivePrefix = "arXiv",
    primaryClass = "astro-ph.CO",
    reportNumber = "IPPP-12-44, DCPT-12-88",
    doi = "10.1088/1475-7516/2012/12/027",
    journal = "JCAP",
    volume = "12",
    pages = "027",
    year = "2012"
}

@article{Ho:2012ug,
    author = "Ho, Chiu Man and Scherrer, Robert J.",
    title = "{Limits on MeV Dark Matter from the Effective Number of Neutrinos}",
    eprint = "1208.4347",
    archivePrefix = "arXiv",
    primaryClass = "astro-ph.CO",
    doi = "10.1103/PhysRevD.87.023505",
    journal = "Phys. Rev. D",
    volume = "87",
    number = "2",
    pages = "023505",
    year = "2013"
}

@article{Steigman:2013yua,
    author = "Steigman, Gary",
    title = "{Equivalent Neutrinos, Light WIMPs, and the Chimera of Dark Radiation}",
    eprint = "1303.0049",
    archivePrefix = "arXiv",
    primaryClass = "astro-ph.CO",
    doi = "10.1103/PhysRevD.87.103517",
    journal = "Phys. Rev. D",
    volume = "87",
    number = "10",
    pages = "103517",
    year = "2013"
}

@article{Nollett:2013pwa,
    author = "Nollett, Kenneth M. and Steigman, Gary",
    title = "{BBN And The CMB Constrain Light, Electromagnetically Coupled WIMPs}",
    eprint = "1312.5725",
    archivePrefix = "arXiv",
    primaryClass = "astro-ph.CO",
    doi = "10.1103/PhysRevD.89.083508",
    journal = "Phys. Rev. D",
    volume = "89",
    number = "8",
    pages = "083508",
    year = "2014"
}

@article{Steigman:2014pfa,
    author = "Steigman, Gary and Nollett, Kenneth M.",
    title = "{Light WIMPs, Equivalent Neutrinos, BBN, and the CMB}",
    eprint = "1401.5488",
    archivePrefix = "arXiv",
    primaryClass = "astro-ph.CO",
    journal = "Mem. Soc. Ast. It.",
    volume = "85",
    pages = "175",
    year = "2014"
}

@article{Nollett:2014lwa,
    author = "Nollett, Kenneth M. and Steigman, Gary",
    title = "{BBN And The CMB Constrain Neutrino Coupled Light WIMPs}",
    eprint = "1411.6005",
    archivePrefix = "arXiv",
    primaryClass = "astro-ph.CO",
    doi = "10.1103/PhysRevD.91.083505",
    journal = "Phys. Rev. D",
    volume = "91",
    number = "8",
    pages = "083505",
    year = "2015"
}

@article{Giacchino:2013bta,
    author = "Giacchino, Federica and Lopez-Honorez, Laura and Tytgat, Michel H. G.",
    title = "{Scalar Dark Matter Models with Significant Internal Bremsstrahlung}",
    eprint = "1307.6480",
    archivePrefix = "arXiv",
    primaryClass = "hep-ph",
    reportNumber = "ULB-TH-13-09",
    doi = "10.1088/1475-7516/2013/10/025",
    journal = "JCAP",
    volume = "10",
    pages = "025",
    year = "2013"
}

@article{Camarena:2024daj,
    author = "Camarena, David and Cyr-Racine, Francis-Yan",
    title = "{Strong constraints on a simple self-interacting neutrino cosmology}",
    eprint = "2403.05496",
    archivePrefix = "arXiv",
    primaryClass = "astro-ph.CO",
    doi = "10.1103/PhysRevD.111.023504",
    journal = "Phys. Rev. D",
    volume = "111",
    number = "2",
    pages = "023504",
    year = "2025"
}

@article{Wilkinson:2014ksa,
    author = "Wilkinson, Ryan J. and Boehm, Celine and Lesgourgues, Julien",
    title = "{Constraining Dark Matter-Neutrino Interactions using the CMB and Large-Scale Structure}",
    eprint = "1401.7597",
    archivePrefix = "arXiv",
    primaryClass = "astro-ph.CO",
    reportNumber = "IPPP-14-03, DCPT-14-06, CERN-PH-TH-2014-013, LAPTH-006-14",
    doi = "10.1088/1475-7516/2014/05/011",
    journal = "JCAP",
    volume = "05",
    pages = "011",
    year = "2014"
}

@article{Crumrine:2024sdn,
    author = "Crumrine, Wendy and Nadler, Ethan O. and An, Rui and Gluscevic, Vera",
    title = "{Dark matter coupled to radiation: Limits from the Milky~Way satellites}",
    eprint = "2406.19458",
    archivePrefix = "arXiv",
    primaryClass = "astro-ph.CO",
    doi = "10.1103/PhysRevD.111.023530",
    journal = "Phys. Rev. D",
    volume = "111",
    number = "2",
    pages = "023530",
    year = "2025"
}

@article{Giare:2023qqn,
    author = "Giar{\`e}, William and G{\'o}mez-Valent, Adri{\`a} and Di Valentino, Eleonora and van de Bruck, Carsten",
    title = "{Hints of neutrino dark matter scattering in the CMB? Constraints from the marginalized and profile distributions}",
    eprint = "2311.09116",
    archivePrefix = "arXiv",
    primaryClass = "astro-ph.CO",
    doi = "10.1103/PhysRevD.109.063516",
    journal = "Phys. Rev. D",
    volume = "109",
    number = "6",
    pages = "063516",
    year = "2024"
}

@article{Pasquini:2015fjv,
    author = "Pasquini, P. S. and Peres, O. L. G.",
    title = "{Bounds on Neutrino-Scalar Yukawa Coupling}",
    eprint = "1511.01811",
    archivePrefix = "arXiv",
    primaryClass = "hep-ph",
    doi = "10.1103/PhysRevD.93.053007",
    journal = "Phys. Rev. D",
    volume = "93",
    number = "5",
    pages = "053007",
    year = "2016",
    note = "[Erratum: Phys.Rev.D 93, 079902 (2016)]"
}

@article{Berryman:2018ogk,
    author = "Berryman, Jeffrey M. and De Gouv{\^e}a, Andr{\'e} and Kelly, Kevin J. and Zhang, Yue",
    title = "{Lepton-Number-Charged Scalars and Neutrino Beamstrahlung}",
    eprint = "1802.00009",
    archivePrefix = "arXiv",
    primaryClass = "hep-ph",
    reportNumber = "NUHEP-TH-18-03, FERMILAB-PUB-18-020-T",
    doi = "10.1103/PhysRevD.97.075030",
    journal = "Phys. Rev. D",
    volume = "97",
    number = "7",
    pages = "075030",
    year = "2018"
}

@article{Bolton:2020ncv,
    author = "Bolton, Patrick D. and Deppisch, Frank F. and Gr{\'a}f, Luk{\'a}{\v{s}} and {\v{S}}imkovic, Fedor",
    title = "{Two-Neutrino Double Beta Decay with Sterile Neutrinos}",
    eprint = "2011.13387",
    archivePrefix = "arXiv",
    primaryClass = "hep-ph",
    doi = "10.1103/PhysRevD.103.055019",
    journal = "Phys. Rev. D",
    volume = "103",
    number = "5",
    pages = "055019",
    year = "2021"
}

@article{CUPID-0:2022yws,
    author = "Azzolini, O. and others",
    collaboration = "CUPID-0",
    title = "{Search for Majoron-like particles with CUPID-0}",
    eprint = "2209.09490",
    archivePrefix = "arXiv",
    primaryClass = "hep-ex",
    doi = "10.1103/PhysRevD.107.032006",
    journal = "Phys. Rev. D",
    volume = "107",
    number = "3",
    pages = "032006",
    year = "2023"
}

@article{Kharusi:2021jez,
    author = "Kharusi, S. Al and others",
    title = "{Search for Majoron-emitting modes of $^{136}$Xe double beta decay with the complete EXO-200 dataset}",
    eprint = "2109.01327",
    archivePrefix = "arXiv",
    primaryClass = "hep-ex",
    doi = "10.1103/PhysRevD.104.112002",
    journal = "Phys. Rev. D",
    volume = "104",
    number = "11",
    pages = "112002",
    year = "2021"
}

@article{KamLAND-Zen:2012uen,
    author = "Gando, A. and others",
    collaboration = "KamLAND-Zen",
    title = "{Limits on Majoron-emitting double-beta decays of Xe-136 in the KamLAND-Zen experiment}",
    eprint = "1205.6372",
    archivePrefix = "arXiv",
    primaryClass = "hep-ex",
    doi = "10.1103/PhysRevC.86.021601",
    journal = "Phys. Rev. C",
    volume = "86",
    pages = "021601",
    year = "2012"
}

@article{Planck:2018vyg,
    author = "Aghanim, N. and others",
    collaboration = "Planck",
    title = "{Planck 2018 results. VI. Cosmological parameters}",
    eprint = "1807.06209",
    archivePrefix = "arXiv",
    primaryClass = "astro-ph.CO",
    doi = "10.1051/0004-6361/201833910",
    journal = "Astron. Astrophys.",
    volume = "641",
    pages = "A6",
    year = "2020",
    note = "[Erratum: Astron.Astrophys. 652, C4 (2021)]"
}

@book{Kolb:1990vq,
    author = "Kolb, Edward W. and Turner, Michael S.",
    title = "{The Early Universe}",
    reportNumber = "FERMILAB-BOOK-1990-01",
    doi = "10.1201/9780429492860",
    isbn = "978-0-429-49286-0, 978-0-201-62674-2",
    publisher = "Taylor and Francis",
    volume = "69",
    month = "5",
    year = "2019"
}

@article{Scherrer:1985zt,
    author = "Scherrer, Robert J. and Turner, Michael S.",
    title = "{On the Relic, Cosmic Abundance of Stable Weakly Interacting Massive Particles}",
    reportNumber = "FERMILAB-PUB-85-163-A, EFI-85-76-CHICAGO",
    doi = "10.1103/PhysRevD.33.1585",
    journal = "Phys. Rev. D",
    volume = "33",
    pages = "1585",
    year = "1986",
    note = "[Erratum: Phys.Rev.D 34, 3263 (1986)]"
}

@article{Wilks:1938dza,
	author = "Wilks, S.S.",
	title = "{The Large-Sample Distribution of the Likelihood Ratio for Testing Composite Hypotheses}",
	doi = "10.1214/aoms/1177732360",
	journal = "Annals Math. Statist.",
	volume = "9",
	number = "1",
	pages = "60--62",
	year = "1938"
}

@article{Cowan:2010js,
	author = "Cowan, Glen and Cranmer, Kyle and Gross, Eilam and Vitells, Ofer",
	title = "{Asymptotic formulae for likelihood-based tests of new physics}",
	eprint = "1007.1727",
	archivePrefix = "arXiv",
	primaryClass = "physics.data-an",
	doi = "10.1140/epjc/s10052-011-1554-0",
	journal = "Eur. Phys. J. C",
	volume = "71",
	pages = "1554",
	year = "2011",
	note = "[Erratum: Eur.Phys.J.C 73, 2501 (2013)]"
}

@article{Burns:2011xf,
	author = "Burns, Eric and Fisher, Wade",
	title = "{Testing the approximations described in 'Asymptotic formulae for likelihood-based tests of new physics'}",
	eprint = "1110.5002",
	archivePrefix = "arXiv",
	primaryClass = "hep-ex",
	month = "10",
	year = "2011"
}

@article{Tanabashi:2018oca,
      author         = "Tanabashi, M. and others",
      title          = "{Review of Particle Physics}",
      collaboration  = "Particle Data Group",
      journal        = "Phys. Rev.",
      volume         = "D98",
      year           = "2018",
      number         = "3",
      pages          = "030001",
      doi            = "10.1103/PhysRevD.98.030001",
      SLACcitation   = "%%CITATION = PHRVA,D98,030001;%%"
}

@article{GERDA:2020xhi,
    author = "Agostini, M. and others",
    collaboration = "GERDA",
    title = "{Final Results of GERDA on the Search for Neutrinoless Double-$\beta$ Decay}",
    eprint = "2009.06079",
    archivePrefix = "arXiv",
    primaryClass = "nucl-ex",
    doi = "10.1103/PhysRevLett.125.252502",
    journal = "Phys. Rev. Lett.",
    volume = "125",
    number = "25",
    pages = "252502",
    year = "2020"
}

@article{Zsigmond:2020bfx,
	author = "Zsigmond, Anna Julia",
	editor = "Nakahata, Masayuki",
	collaboration = "LEGEND",
	title = "{LEGEND: The future of neutrinoless double-beta decay search with germanium detectors}",
	doi = "10.1088/1742-6596/1468/1/012111",
	journal = "J. Phys. Conf. Ser.",
	volume = "1468",
	number = "1",
	pages = "012111",
	year = "2020"
}

@article{CUPIDInterestGroup:2019inu,
	author = "Armstrong, W.R. and others",
	collaboration = "CUPID",
	title = "{CUPID pre-CDR}",
	eprint = "1907.09376",
	archivePrefix = "arXiv",
	primaryClass = "physics.ins-det",
	month = "7",
	year = "2019"
}

@article{NEMO-3:2019gwo,
	author = "Arnold, R. and others",
	collaboration = "NEMO-3",
	title = "{Detailed studies of $^{100}$Mo two-neutrino double beta decay in NEMO-3}",
	eprint = "1903.08084",
	archivePrefix = "arXiv",
	primaryClass = "nucl-ex",
	doi = "10.1140/epjc/s10052-019-6948-4",
	journal = "Eur. Phys. J. C",
	volume = "79",
	number = "5",
	pages = "440",
	year = "2019"
}

@article{KamLAND-Zen:2019imh,
	author = "Gando, A. and others",
	collaboration = "KamLAND-Zen",
	title = "{Precision measurement of the $^{136}$Xe two-neutrino $\beta\beta$ spectrum in KamLAND-Zen and its impact on the quenching of nuclear matrix elements}",
	eprint = "1901.03871",
	archivePrefix = "arXiv",
	primaryClass = "hep-ex",
	reportNumber = "INT-PUB-19-001",
	doi = "10.1103/PhysRevLett.122.192501",
	journal = "Phys. Rev. Lett.",
	volume = "122",
	number = "19",
	pages = "192501",
	year = "2019"
}

@article{Jokiniemi:2022ayc,
    author = "Jokiniemi, Lotta and Romeo, Beatriz and Soriano, Pablo and Men{\'e}ndez, Javier",
    title = "{Neutrinoless {\ensuremath{\beta}}{\ensuremath{\beta}}-decay nuclear matrix elements from two-neutrino {\ensuremath{\beta}}{\ensuremath{\beta}}-decay data}",
    eprint = "2207.05108",
    archivePrefix = "arXiv",
    primaryClass = "nucl-th",
    doi = "10.1103/PhysRevC.107.044305",
    journal = "Phys. Rev. C",
    volume = "107",
    number = "4",
    pages = "044305",
    year = "2023"
}

@article{nEXO:2021ujk,
    author = "Adhikari, G. and others",
    collaboration = "nEXO",
    title = "{nEXO: neutrinoless double beta decay search beyond 10$^{28}$ year half-life sensitivity}",
    eprint = "2106.16243",
    archivePrefix = "arXiv",
    primaryClass = "nucl-ex",
    doi = "10.1088/1361-6471/ac3631",
    journal = "J. Phys. G",
    volume = "49",
    number = "1",
    pages = "015104",
    year = "2022"
}

@article{Doi:1985dx,
    author = "Doi, M. and Kotani, T. and Takasugi, E.",
    title = "{Double beta Decay and Majorana Neutrino}",
    reportNumber = "OS-GE-85-03",
    doi = "10.1143/PTPS.83.1",
    journal = "Prog. Theor. Phys. Suppl.",
    volume = "83",
    pages = "1",
    year = "1985"
}

@article{PandaX:2025tls,
    author = "Li, Tao and others",
    collaboration = "PandaX",
    title = "{Probing scalar-neutrino and scalar-dark-matter interactions with PandaX-4T}",
    eprint = "2511.13515",
    archivePrefix = "arXiv",
    primaryClass = "hep-ex",
    month = "11",
    year = "2025"
}

@article{Frigerio:2011in,
    author = "Frigerio, Michele and Hambye, Thomas and Masso, Eduard",
    title = "{Sub-GeV dark matter as pseudo-Goldstone from the seesaw scale}",
    eprint = "1107.4564",
    archivePrefix = "arXiv",
    primaryClass = "hep-ph",
    reportNumber = "ULB-TH-11-17",
    doi = "10.1103/PhysRevX.1.021026",
    journal = "Phys. Rev. X",
    volume = "1",
    pages = "021026",
    year = "2011"
}

@article{Burgess:1992dt,
    author = "Burgess, C. P. and Cline, James M.",
    title = "{Majorons without Majorana masses and neutrinoless double beta decay}",
    eprint = "hep-ph/9209299",
    archivePrefix = "arXiv",
    reportNumber = "MCGILL-92-22",
    doi = "10.1016/0370-2693(93)91720-8",
    journal = "Phys. Lett. B",
    volume = "298",
    pages = "141--148",
    year = "1993"
}

@article{Burgess:1993xh,
    author = "Burgess, C. P. and Cline, James M.",
    title = "{A New class of Majoron emitting double beta decays}",
    eprint = "hep-ph/9307316",
    archivePrefix = "arXiv",
    reportNumber = "MCGILL-93-02, TPI-MINN-93-28-T, UMN-TH-1204-93",
    doi = "10.1103/PhysRevD.49.5925",
    journal = "Phys. Rev. D",
    volume = "49",
    pages = "5925--5944",
    year = "1994"
}

@article{Agostini:2020cpz,
    author = "Agostini, Matteo and Bossio, Elisabetta and Ibarra, Alejandro and Marcano, Xabier",
    title = "{Search for Light Exotic Fermions in Double-Beta Decays}",
    eprint = "2012.09281",
    archivePrefix = "arXiv",
    primaryClass = "hep-ph",
    reportNumber = "TUM-HEP 1306/20",
    doi = "10.1016/j.physletb.2021.136127",
    journal = "Phys. Lett. B",
    volume = "815",
    pages = "136127",
    year = "2021"
}

@article{Nozzoli:2022tov,
    author = "Nozzoli, Francesco and Cernetti, Cinzia",
    title = "{Dark Matter stimulated neutrinoless double beta decay}",
    eprint = "2212.07832",
    archivePrefix = "arXiv",
    primaryClass = "hep-ph",
    month = "12",
    year = "2022"
}

@article{Kamiokande-II:1987idp,
    author = "Hirata, K. and others",
    editor = "Wali, K. C.",
    collaboration = "Kamiokande-II",
    title = "{Observation of a Neutrino Burst from the Supernova SN 1987a}",
    reportNumber = "UT-ICEPP-87-01, UPR-142E",
    doi = "10.1103/PhysRevLett.58.1490",
    journal = "Phys. Rev. Lett.",
    volume = "58",
    pages = "1490--1493",
    year = "1987"
}

@article{Hirata:1988ad,
    author = "Hirata, K. S. and others",
    title = "{Observation in the Kamiokande-II Detector of the Neutrino Burst from Supernova SN 1987a}",
    doi = "10.1103/PhysRevD.38.448",
    journal = "Phys. Rev. D",
    volume = "38",
    pages = "448--458",
    year = "1988"
}

@article{Alekseev:1988gp,
    author = "Alekseev, E. N. and Alekseeva, L. N. and Krivosheina, I. V. and Volchenko, V. I.",
    title = "{Detection of the Neutrino Signal From {SN1987A} in the {LMC} Using the Inr Baksan Underground Scintillation Telescope}",
    doi = "10.1016/0370-2693(88)91651-6",
    journal = "Phys. Lett. B",
    volume = "205",
    pages = "209--214",
    year = "1988"
}

@article{Alekseev:1987ej,
    author = "Alekseev, E. N. and Alekseeva, L. N. and Volchenko, V. I. and Krivosheina, I. V.",
    editor = "Tran Thanh Van, J.",
    title = "{Possible Detection of a Neutrino Signal on 23 February 1987 at the Baksan Underground Scintillation Telescope of the Institute of Nuclear Research}",
    journal = "JETP Lett.",
    volume = "45",
    pages = "589--592",
    year = "1987"
}

@article{Bionta:1987qt,
    author = "Bionta, R. M. and others",
    title = "{Observation of a Neutrino Burst in Coincidence with Supernova SN 1987a in the Large Magellanic Cloud}",
    reportNumber = "UCI-NEUTRINO-87-10",
    doi = "10.1103/PhysRevLett.58.1494",
    journal = "Phys. Rev. Lett.",
    volume = "58",
    pages = "1494",
    year = "1987"
}

@article{Blinov:2019gcj,
    author = "Blinov, Nikita and Kelly, Kevin James and Krnjaic, Gordan Z and McDermott, Samuel D",
    title = "{Constraining the Self-Interacting Neutrino Interpretation of the Hubble Tension}",
    eprint = "1905.02727",
    archivePrefix = "arXiv",
    primaryClass = "astro-ph.CO",
    reportNumber = "FERMILAB-PUB-19-175-A-T",
    doi = "10.1103/PhysRevLett.123.191102",
    journal = "Phys. Rev. Lett.",
    volume = "123",
    number = "19",
    pages = "191102",
    year = "2019"
}

@article{Deppisch:2020ztt,
    author = "Deppisch, Frank F. and Graf, Lukas and Iachello, Francesco and Kotila, Jenni",
    title = "{Analysis of light neutrino exchange and short-range mechanisms in $0\nu\beta\beta$ decay}",
    eprint = "2009.10119",
    archivePrefix = "arXiv",
    primaryClass = "hep-ph",
    doi = "10.1103/PhysRevD.102.095016",
    journal = "Phys. Rev. D",
    volume = "102",
    number = "9",
    pages = "095016",
    year = "2020"
}

@article{Huang:2014bva,
    author = "Huang, Wei-Chih and Deppisch, Frank F.",
    title = "{Dark matter origins of neutrino masses}",
    eprint = "1412.2027",
    archivePrefix = "arXiv",
    primaryClass = "hep-ph",
    reportNumber = "LCTS-2014-48",
    doi = "10.1103/PhysRevD.91.093011",
    journal = "Phys. Rev. D",
    volume = "91",
    pages = "093011",
    year = "2015"
}

@article{Telalovic:2024cot,
    author = "Telalovic, Bernanda and Fiorillo, Damiano F. G. and Mart{\'\i}nez-Mirav{\'e}, Pablo and Vitagliano, Edoardo and Bustamante, Mauricio",
    title = "{The next galactic supernova can uncover mass and couplings of particles decaying to neutrinos}",
    eprint = "2406.15506",
    archivePrefix = "arXiv",
    primaryClass = "hep-ph",
    doi = "10.1088/1475-7516/2024/11/011",
    journal = "JCAP",
    volume = "11",
    pages = "011",
    year = "2024"
}
\end{document}